\documentclass[11pt, a4paper]{article}
\usepackage[top=1in, bottom=1in, left=1.25in, right=1.25in]{geometry}
\usepackage{booktabs}
\usepackage{multirow}
\usepackage{setspace}
\usepackage{natbib}
\usepackage{authblk}
\usepackage[pdftex]{graphicx}
\usepackage{amsmath}
\usepackage{amsthm}
\usepackage{mathtools}
\DeclarePairedDelimiter\set{\{}{\}}
\DeclarePairedDelimiterX\Set[2]{\{}{\}}{\mspace{2mu}{#1}\;\delimsize|\;{#2}\mspace{2mu}}
\mathtoolsset{showonlyrefs}
\usepackage{amsfonts}
\usepackage{ascmac}
\usepackage{amssymb}
\usepackage{bm}
\usepackage{bbm}
\usepackage{algorithmic}
\usepackage{algorithm}
\usepackage{color}
\usepackage{enumitem}
\usepackage{mathrsfs}
\usepackage[, 
	colorlinks = true, 
	anchorcolor=black,
	linkcolor = blue,
	citecolor = blue, 
	filecolor  = blue, 
	urlcolor   = blue
]{hyperref}
\usepackage{comment}
\usepackage[toc,page,titletoc]{appendix}

\theoremstyle{plain}
\newtheorem{theorem}{Theorem}[section]

\theoremstyle{definition}
\newtheorem{definition}[theorem]{Definition}

\theoremstyle{remark}
\newtheorem{remark}{Remark}

\newcommand{\ep}{\varepsilon}

\newcommand{\E}{\mathbb{E}}

\newcommand{\mcal}[1]{\mathcal{#1}}

\def\dd{\textrm{d}}

\newcommand{\argmin}{\operatornamewithlimits{argmin}}

\title{\bf Robust estimation via $\gamma$-divergence for diffusion processes}
\author[1,2]{Tomoyuki Nakagawa}
\author[3]{Yusuke Shimizu}

\affil[1]{School of Data Science, Meisei University\footnote{2-1-1 Hodokubo, Hino, Tokyo 191-8506, Japan}}
\affil[2]{Statistical Mathematics Unit, RIKEN Center for Brain Science\footnote{2-1 Hirosawa Wako City, Saitama 351-0198 Japan}}
\affil[3]{Department of Mathematics, Josai University\footnote{Keyakidai 1-1, Sakado-city, Saitama 350-0246, Japan}}
\date{Last update: \today}

\begin{document}
\maketitle
\begin{abstract}
This paper deals with the problem of outliers in high frequency observation data from diffusion processes. 
    Robust estimation methods are needed because the inclusion of outliers can lead to incorrect statistical inference even in the diffusion process. 
    To construct a robust estimator, we first approximate the transition density of the diffusion process to the Gaussian density by using Kessler's approach and then employ two types of minimum robust divergence estimation methods. 
    In this paper, we provide the asymptotic properties of the robust estimator using $\gamma$-divergence. 
  Furthermore, we derive the conditional influence functions of the estimation using divergences and discuss its boundness.

\end{abstract}


\medskip

\section{Introduction}

This paper deals with the problem of outliers in high frequency observation data from diffusion processes. 
Diffusion processes are popular in fields such as the physical and biological sciences, finance, and engineering.
Over the past few decades, many papers have been devoted to statistical inference for diffusion processes; see \cite{kessler1997estimation, masuda2017moment, uchida2012adaptive, yoshida2011polynomial} and references therein.
Especially, estimation in discretely observed diffusion processes has received a great deal of attention. In particular, Kessler \cite{kessler1997estimation} presented the quasi-likelihood function of a diffusion process and derived the asymptotic distribution of the minimum contrast estimator. However, it is well-known that estimators based on likelihood are strongly influenced by outliers or extreme values. 
To overcome this problem, various robust estimation methods have been developed; see \cite{la2010infinitesimal, yoshida1988robust} and references therein.
Among them, we especially consider robust estimation using the density power divergence, proposed by \cite{basu1998robust}, and $\gamma$-divergence, proposed by \cite{jones2001comparison}, for diffusion processes in this paper.  
The robustness against outliers has been investigated in many aspects such as influence function, breakdown point, and redescending (for details, see \cite{hampel1986robust, huber2009robust, maronna2019robust} etc.). 
Recently, \cite{lee2013minimum} proved the asymptotic properties of the robust estimator using the density power divergence on the Ornstein-Uhlenbeck type processes. 
Similarly,  \cite{song2007minimum} and \cite{song2017robust} proved the asymptotic properties of the robust estimator using the density power divergence of the dispersion parameter in discretely observed diffusion processes. 
However, they make little mention of the robustness of the density power divergence-based estimator.

The purpose of this paper is to propose a robust estimator using divergence in discretely observed diffusion processes. In particular, we show the robustness (e.g. the bounded influence and the redescending, etc.) of two robust estimators using the density power- and $\gamma$-divergences. 
This paper is organized as follows: Section \ref{sec:divergence} introduces the density power- and $\gamma$-divergences-based estimation in discretely observed diffusion processes. 
Section \ref{sec:asymp} provides the consistency and the asymptotic normality of the proposed estimators.
In Section \ref{sec:influence}, we derive the conditional influence function (defined by \cite{la2010infinitesimal}) of the density power- and $\gamma$-divergences-based estimation, and show the robustness of them. 
In Section \ref{sec:sim}, we consider several probability models for time series outliers, including additive outliers (AOs) and replacement outliers (ROs) as described in \cite{maronna2019robust}, and compare the robust divergence-based estimation with the likelihood-based estimation using the Monte Carlo method.

\section{Robust Divergence estimation}
\label{sec:divergence}
Consider the estimation method based on divergence that evaluates the discrepancy between any two probability distributions. 
The divergence-based estimation methods have been used successfully in constructing robust estimators. 
 For a review, we refer to \cite{pardo2006statistical}, \cite{cichocki2010families}, and the references therein.
In particular, the density power divergence and the $\gamma$-divergence are well-known as robust divergence and have good robust properties (e.g. the bounded influence and the redescending, etc.) in kinds of literature (see \cite{fujisawa2008robust, momozaki2022robustness} etc.).

\subsection{Robust divergence}
In this section, we introduce the density power divergence and the $\gamma$-divergence. 
Suppose that $f$ and $g$ are the probability density functions. 

\cite{basu1998robust} proposed the density power divergence defined as
\begin{align}
\begin{split}\label{siki1}
    \mcal{D}_{\alpha}(g, f) &= \frac{1}{\alpha(\alpha + 1)}\int_{\Omega} g(x)^{\alpha + 1}\dd x -\frac{1}{\alpha}\int_{\Omega} g(x) f(x)^{\alpha}\dd x\\
& \hspace{10pt} + \frac{1}{(1+ \alpha)}\int_{\Omega} f(x)^{1+\alpha}\dd x, \quad (\alpha>0). 
\end{split}
\end{align}
This divergence includes Kullback-Leibler (KL) divergence and $L_2$-distance as special cases.
\cite{jones2001comparison} also proposed the $\gamma$-divergence defined as 
\begin{align}
\begin{split}\label{siki2}
\mcal{D}_{\gamma}(g, f) &= \frac{1}{\gamma(\gamma + 1)}\int_{\Omega} g(x)^{\gamma + 1}\dd x -\frac{1}{\gamma}\log\int_{\Omega}g(x)f(x)^{\gamma}\dd x\\
& \hspace{10pt} + \frac{1}{(1+ \gamma)}\log\int_{\Omega} f(x)^{1+\gamma} \dd x, \quad (\gamma>0).
\end{split}
\end{align}
Here, $\gamma$-divergence includes KL divergence. 

For a family of parametric distributions $\Set*{F_{\theta}}{\theta \in \Theta}$ possessing densities $\set{f_{\theta}}$ and for a distribution $G$ with density $g$, we define the minimum density power- and $\gamma$- divergence functional $T_{\alpha}(\cdot)$ and $T_{\gamma}(\cdot)$ by 
\begin{align*}
    T_{\alpha}(g) = \argmin_{\theta \in \Theta} \mcal{D}_{\alpha}(g, f_{\theta}), \quad T_{\gamma}(g) = \argmin_{\theta \in \Theta} \mcal{D}_{\gamma}(g, f_{\theta}). 
\end{align*}
The minimum density power- and $\gamma$- divergence estimators by minimizing the empirical version of the density power- and $\gamma$- divergence are weakly consistent for $T_{\alpha}(g)$ and $T_{\gamma}(g)$, and asymptotically normal, and demonstrated that the estimator has strong robust properties against outliers and the misspecification of underlying models (for details, see \cite{basu1998robust, jones2001comparison, fujisawa2008robust} etc.). 

These approaches can be extended to regression models (for details, see \cite{basu1998robust, fujisawa2008robust} and references therein.).  
Let $\set*{f_{\theta}(y \mid x)}$ be a
parametric family of regression models indexed by the parameter $\theta \in \Theta$
and let $g(y\mid x)$ be the true density for $Y$ given $X = x$. Substituting $f$ and $g$ in \eqref{siki1} by $f_{\theta}(y\mid x)$ and $g(y\mid x)$ respectively, a family of the $x$-conditional versions of the density power divergence is obtained as follows:
\begin{align}
\begin{split}\label{siki3}
    &\mcal{D}_{\alpha}\left(g(\cdot \mid x), f(\cdot \mid x)\right) = \frac{1}{\alpha(\alpha + 1)}\int_{\Omega} g(y\mid x)^{\alpha + 1}\dd y-\frac{1}{\alpha}\int_{\Omega} g(y\mid x) f(y\mid x)^{\alpha}\dd y \\
& \hspace{100pt} + \frac{1}{(1+ \alpha)}\int_{\Omega} f(y\mid x)^{1+\alpha}\dd y, \quad (\alpha>0). 
\end{split}
\end{align}
Similarly, substituting $f$ and $g$ in \eqref{siki2} by $f_{\theta}(y\mid x)$ and $g(y\mid x)$ respectively, a family of the $x$-conditional versions of $\gamma$-divergence is obtained as follows:
\begin{align}
\begin{split}\label{siki4}
&\mcal{D}_{\gamma}\left(g(\cdot \mid x), f(\cdot \mid x)\right) = \frac{1}{\gamma(\gamma + 1)}\log\int_{\Omega} g(y\mid x)^{\gamma + 1}\dd y -\frac{1}{\gamma}\log\int_{\Omega}g(y\mid x)f_{\theta}(y\mid x)^{\gamma}\dd y\\
& \hspace{100pt} + \frac{1}{(1+ \gamma)}\log\int_{\Omega} f_{\theta}(y\mid x)^{1+\gamma} \dd y, \quad (\gamma>0).
\end{split}
\end{align}
Using \eqref{siki3} and \eqref{siki4}, we consider extending these approaches to the diffusion process in the following section. 
\subsection{Estimation for SDE using Robust divergence}
Consider a one-dimensional ergodic diffusion process, which is described by Wiener-driven stochastic differential equation (SDE):
\begin{align}
dX_{t}=b(X_{t},\mu)dt+a(X_{t},\sigma)dw_{t},\quad X_{0}=x_{0}, \label{diff}
\end{align}
where $\{w_{t}\}_{t \geq 0}$ is a one-dimensional Wiener process, $x_{0}$ is a random variable independent of $\{w_{t}\}_{t\geq0}$, $\sigma\in\Theta_{\sigma}$ and $\mu \in\Theta_{\mu}$ are unknown parameters, $\Theta_{\sigma}\subset\mathbb{R}^{p}$ and $\Theta_{\mu}\subset\mathbb{R}^{q}$ are compact sets, and
$a:\mathbb{R}\times\mathbb{R}^{p}\to\mathbb{R}$ and $b:\mathbb{R}\times\mathbb{R}^{q}\to\mathbb{R}$ are known functions. $\theta_{0}:=(\mu_{0}, \sigma_{0})$ is the true value of $\theta:=(\mu, \sigma)$.
The data are discrete observations $\{X_{t_{i}^{n}}\}_{i=0,1,\ldots,n}$, where $t_{i}^{n}=ih_{n}$ ($h_{n}>0$ is the discretisation step).
We will consider the situation when $h_{n}\to0$, $nh_{n}\to\infty$ and $nh_{n}^{2}\to 0 \ (n\to\infty)$.

\cite{lee2013minimum} and \cite{song2017robust} give the density power cross entropy for the diffusion process as follows: 
\begin{align}
Q_{n, \alpha}(\theta)&= \sum_{i=1}^{n}\left\{-\frac{1}{\alpha}f_{\theta}(x_{t_i^n} \mid x_{t_{i-1}^{n}})^{\alpha} + \frac{1}{\alpha + 1}\int f_{\theta}(x \mid x_{t_{i-1}^{n}})^{1+\alpha} \dd x \right\}=\sum_{i = 1}^n q_{\alpha, i}(\theta),
\end{align}
where
\begin{align*}
 q_{\alpha, i}(\theta) = \begin{dcases}
 \begin{array}{c}
 \displaystyle \frac{1}{(2\pi a_{i-1}(\sigma)^2h_n)^{\alpha/2}}\left[ -\frac{1}{\alpha}\exp\left\{- \frac{\alpha(X_{t_{i}^n} - X_{t_{i-1}^n} - h_n b_{i-1}(\mu))^2}{2a_{i-1}(\sigma)^2 h_n}\right\}\right.\\
 \displaystyle \left. \hspace{30pt}+ \frac{1}{(1 + \alpha)^{3/2}}\right]
 \end{array}, & (\alpha > 0), \\
 \frac{1}{2}\log(2\pi a_{i -1}(\sigma)^2) + \frac{\left(X_{t_{i}^n} - X_{t_{i-1}^n} - h_n b_{i-1}(\mu)\right)^2}{2 a_{i -1}(\sigma)^2 h_n}, & (\alpha = 0), 
 \end{dcases}
\end{align*}
$a_{i -1}(\sigma) = a(X_{t_{i -1}^n}, \sigma)$ and $b_{i -1}(\mu) = b(X_{t_{i -1}^n}, \mu)$. 
Then they proved the consistency and the asymptotic normality of estimators using the density power divergence. 
Moreover, \cite{ghosh2022general} discussed its exponential convergence in Bayesian contexts.

Similarly, we define the $\gamma$-cross entropy for the diffusion process model as follows: 
\begin{align*}
Q_{n, \gamma}(\theta)&= \sum_{i=1}^{n}\frac{-f_{\theta}(X_{t_i^n} \mid X_{t_{i-1}^{n}})^{\gamma}}{\left\{\int f_{\theta}(x \mid X_{t_{i-1}^{n}})^{1+\gamma} \dd x \right\}^{\gamma/(1+ \gamma)}}=\sum_{i = 1}^n q_{\gamma, i}(\theta),
\end{align*}
where 
\begin{align}
 q_{\gamma, i}(\theta) = \begin{dcases}
-\left(\frac{1 + \gamma}{2 \pi a_{i -1}(\sigma)^2h_{n}} \right)^{\gamma/2(1 + \gamma)}\exp\left\{- \frac{\gamma(X_{t_{i}^n} - X_{t_{i-1}^n} - h_n b_{i-1}(\mu))^2}{2a_{i-1}(\sigma)^2 h_n}\right\},& (\gamma > 0),\\
 \frac{1}{2}\log(2\pi a_{i -1}(\sigma)^2) + \frac{\left(X_{t_{i}^n} - X_{t_{i-1}^n} - h_n b_{i-1}(\mu)\right)^2}{2 a_{i -1}(\sigma)^2 h_n}, & (\gamma = 0).
 \end{dcases}
 \label{eq:gamma}
 \end{align}

\noindent It should be noted that $q_{0, i}(\theta) = \lim_{\alpha \to 0}q_{\alpha, i}(\theta) = \lim_{\gamma \to 0}q_{\gamma, i}(\theta)$. 

\section{Asymptotic properties}
\label{sec:asymp}
In this section, we will derive the asymptotic properties of the 
$\gamma$-divergence based estimator. we consider the SDE
\begin{align}
dX_{t}=b(X_{t},\mu)dt+\sigma dw_{t},\quad X_{0}=x_{0},
\label{eq:SDE1}
\end{align}
where $a(x,\sigma)=\sigma$ in \eqref{diff}. 
Then, let us define the estimator $\hat{\theta}_{n}^{(\gamma)}=(\hat{\mu}_{n}^{(\gamma)},\hat{\sigma}_{n}^{(\gamma)})$ of $\theta$ as
\begin{align*}
Q_{n,\gamma}(\hat{\theta}_{n}^{(\gamma)})&=\inf_{\theta\in\Theta }\sum_{i=1}^{n}q_{\gamma,i}(\theta),
\end{align*}
where 
\begin{align*}
 q_{\gamma, i}(\theta) = \begin{dcases}
-\sigma^{-\gamma/(1 + \gamma)}\exp\left\{- \frac{\gamma(X_{t_{i}^n} - X_{t_{i-1}^n} - h_n b_{i-1}(\mu))^2}{2\sigma^2 h_n}\right\},& (\gamma > 0),\\
 \log\sigma^2 + \frac{\left(X_{t_{i}^n} - X_{t_{i-1}^n} - h_n b_{i-1}(\mu)\right)^2}{\sigma^2 h_n}, & (\gamma = 0).
 \end{dcases}
\end{align*}
Note that we omit the constant term from \eqref{eq:gamma}. We set
\begin{align*}
\mathscr{P}= \Set*{f(x,\mu)}{|f|\leq C(1+|x|)^{C}\ {\text{for some }}C},
\end{align*}
where $C$ does not depend on $\theta$. Then, we make the following assumptions:
\begin{enumerate}[label = (A\arabic*), ref = (A\arabic*)]
\item \label{A1} There exists a constant $C>0$ such that for all $x, y\in\mathbb{R}$, $|b(x,\mu_{0})-b(y,\mu_{0})|\leq C|x-y|$.
\item \label{A2} The process $X$ is ergodic for $(\mu_{0},\sigma_{0})$ with its invariant measure $\nu_{0}$ such that $\int x^{k}d\nu_{0}(x)<\infty$ for all $k\geq0.$
\item \label{A3} $\sup_{t}\E[|X_{t}|^{k}]<\infty$ for all $k>0$.
\item \label{A4} The function $b$ is continuously differentiable with respect to $x$ for all $\mu$ and the derivatives belong to $\mathscr{P}$.
\item \label{A5} The function $b$ and all its $x$-derivatives are three times differentiable with respect to $\mu$ for all $x$. Moreover, these derivatives up to the third order with respect to $\mu$ belong to $\mathscr{P}$.
\item \label{A6} If $b(x,\mu)=b(x,\mu_{0})$ for $\nu_{0}$ a.s. all $x$, then $\mu=\mu_{0}$.
\item \label{A7} $S:=\int\partial_{\mu} b(x,\mu_{0})\partial_{\mu^{\top}}b(x,\mu_{0})d\nu_{0}(x)$ is positive definite, where $\partial_{\mu}b=\partial b/\partial\mu$.
\end{enumerate}
The following theorem gives the asymptotic properties of $\hat{\theta}_{n}^{(\gamma)}$.
\begin{theorem}
\label{thm1}
Assume that \ref{A1}--\ref{A7}, $h_{n}\to0$, $nh_{n}\to\infty$ and $nh_{n}^{2}\to0$. Then for any $\gamma\geq0$,
\begin{align*}
\left(\sqrt{nh_{n}}(\hat{\mu}_{n}^{(\gamma)}-\mu_{0}),\sqrt{n}(\hat{\sigma}_{n}^{(\gamma)}-\sigma_{0})\right)\xrightarrow{\mathcal{L}}N_{p+1}(0,\Sigma_{0}^{(\gamma)}),
\end{align*}
where
\begin{align*}
\Sigma_{0}^{(\gamma)}=\sigma_{0}^{2}\left(
\begin{array}{cc}
\left(\frac{1+\gamma}{\sqrt{1+2\gamma}}\right)^{3}S^{-1} & 0 \\
0 & \frac{(1+\gamma)^{3}(3\gamma^{2}+4\gamma+2)}{4(1+2\gamma)^{5/2}}
\end{array}
\right).
\end{align*}
\end{theorem}

\noindent See Appendix \ref{A.1} for the proof of Theorem \ref{thm1}.
\begin{remark}
The SDE \eqref{eq:SDE1} can be extended to the following form:
\begin{align}
dX_{t}=b(X_{t},\mu)dt+\sigma a(X_{t})dw_{t}.
\label{eq:SDE2}
\end{align}
Since from the transformation $Y_{t}=H(X_{t})$, where $H$ satisfies $\partial_{x}H(x)=1/a(x)$, we obtain
\begin{align*}
dY_{t}=M(Y_{t},\mu,\sigma)dt+\sigma dw_{t},
\end{align*}
where
\begin{align*}
M(y,\mu,\sigma)=\frac{b\left(H^{-1}(y),\mu\right)}{a\left(H^{-1}(y)\right)}-\frac{\sigma^{2}}{2}\partial_{x}a\left(H^{-1}(y),\sigma\right),
\end{align*}
and subsequently, the estimation procedure for \eqref{eq:SDE2} is reduced to that for the SDE \eqref{eq:SDE1}.

\end{remark}

\section{Robustness}
\label{sec:influence}
In this section, we consider infinitesimally robust statistical procedures for the diffusion processes.
The need for robust statistical methodologies is widely recognized in the statistical literature. 
Important contributions studied infinitesimally robust, i.e., bounded-influence and estimators for independent and identically distributed data; see, among others, \cite{hampel1986robust, huber2009robust}.

\cite{la2010infinitesimal} defined the influence function and the conditional influence function for the diffusion process. 
Let 
\begin{align*}
G_{\ep, \nu} = (1 - \ep )G_{\theta_0} + \ep \nu
\end{align*}
with $\nu$ is the contamination distribution, be a generic $\ep$-contamination of the parametric diffusion probability $G_{\theta_0}$, where $0 \leq \ep \leq \eta$ for fixed $0 < \eta < 1$. 
\begin{definition}[Conditional Influence Function]
An influence function (IF) for the conditionally unbiased estimator $\hat{\theta}(\cdot)$ of the diffusion process in \eqref{diff} is any function $\textrm{IF}(\cdot, \cdot; \theta_0)$ such that
\begin{align}
\partial_{\ep}\hat{\theta}(G_{\ep, \nu})\Big\vert_{\ep = 0} = \int \textrm{IF}(x_1, x_0; \theta_0)\nu(\dd x_0, \dd x_1)
\end{align}
for all $\nu \in \mcal{M}$, where $\mcal{M}$ is the family of two-dimensional marginal distributions of a stationary process. 
Moreover, the conditional influence function ($\textrm{IF}^c$) is any influence function $\textrm{IF}^c(\cdot, \cdot; \theta_0)$ such that $\E[\textrm{IF}^c(X_{t_{i}^n}, X_{t_{i-1}^n}; \theta_0)\mid X_{t_{i-1}^n}] = 0$. 
\end{definition}

Moreover, \cite{la2010infinitesimal} proved the following theorem. 

\begin{theorem}[Corollary 5 in \citealp{la2010infinitesimal}]
\label{CIF}
Let $\hat{\theta}$ be a conditionally unbiased $M$-estimator for the diffusion process, defined by an integrable estimating function $\psi$. Then the conditional influence function of $\hat{\theta}$ is uniquely given by
\begin{align}\label{con_influ}
\mathrm{IF}^c_{\psi}(x_i, x_{i-1}; \theta_0) = - D(\psi, \theta_0)^{-1}\psi(x_i, x_{i - 1}; \theta_0)
\end{align}
where $D(\psi, \theta_0)=E_{\theta_{0}}\left[\partial_{\theta} \psi(X_i, X_{i-1}; \theta_0)\right]$. 
\end{theorem}

From Theorem \ref{CIF}, we obtain the conditional influence function of the robust divergences as 
\begin{align*}
\mathrm{IF}^c_{\psi_{\alpha}}(x_i, x_{i-1}; \theta_0) &= -D(\psi_{\alpha}, \theta_0)^{-1} \psi_{\alpha}(x_i, x_{i - 1}; \theta_0), \\
\mathrm{IF}^c_{\psi_{\gamma}}(x_i, x_{i-1}; \theta_0) &=-D(\psi_{\gamma}, \theta_0)^{-1} \psi_{\gamma}(x_i, x_{i - 1}; \theta_0). 
\end{align*}
Here $\psi_{\alpha}$ and $\psi_{\gamma}$ are the estimating functions of the density power- and $\gamma$- divergence respectively, and given as follows:
\begin{align*}
\psi_{\alpha}(x_{i}, x_{i-1}; \theta) &= (\psi_{\alpha, 1}(x_{i}, x_{i-1}; \theta), \psi_{\alpha, 2}(x_{i}, x_{i-1}; \theta))^{\top}, \\
\psi_{\gamma}(x_{i}, x_{i-1}; \theta) &= (\psi_{\gamma, 1}(x_{i}, x_{i-1}; \theta), \psi_{\gamma, 2}(x_{i}, x_{i-1}; \theta))^{\top}, 
\end{align*}
where  
\begin{align*}
  \psi_{\alpha, 1}(x_{i}, x_{i-1}; \theta) &= \partial_{\mu} q_{\alpha,i}(\theta)= -w_i^{(\alpha)} \frac{(x_{i} - x_{i-1} - h_n b_{i-1}(\mu))}{a_{i-1}(\sigma)^2} \partial_{\mu}b_{i-1}(\mu), \\
  \psi_{\alpha, 2}(x_{i}, x_{i-1}; \theta) &=\partial_{\sigma} q_{\alpha,i}(\theta) \\
  &= \left\{ w_i^{(\alpha)} - \frac{\alpha}{(1 + \alpha)^{3/2}}\frac{1}{(2\pi a_{i - 1}(\sigma)^2h_n)^{\alpha/2}} \right\}\frac{1}{a_{i - 1}(\sigma)}\partial_{\sigma}a_{i-1}(\sigma)\\ 
& \hspace{10pt} -  w_i^{(\alpha)}  \frac{(x_{i} - x_{i-1} - h_n b_{i-1}(\mu))^2}{a_{i-1}(\sigma)^3 h_n} \partial_{\sigma}a_{i-1}(\sigma), \\
\psi_{\gamma, 1}(x_{i}, x_{i-1}; \theta) &=\partial_{\mu} q_{\gamma,i}(\theta) =  -w_i^{(\gamma)} \frac{(x_{i} - x_{i-1} - h_n b_{i-1}(\mu))}{a_{i-1}(\sigma)^2}  \partial_{\mu} b_{i-1}(\mu), \\
\psi_{\gamma, 2}(x_{i}, x_{i-1}; \theta) &=\partial_{\sigma} q_{\gamma,i}(\theta) \\
&= \frac{1}{\gamma + 1} w_i^{(\gamma)}  \frac{1}{a_{i -1}(\sigma)}\partial_{\sigma} a_{i-1}(\sigma)\\ 
& \hspace{10pt} -   w_i^{(\gamma)} \frac{(x_{i} - x_{i-1} - h_n b_{i-1}(\mu))^2}{a_{i-1}(\sigma)^3 h_n} \partial_{\sigma} a_{i-1}(\sigma),   
\end{align*}
and 
\begin{align*}
w_i^{(\alpha)} &=  (2 \pi h_n a_{i -1}(\sigma)^2)^{-\alpha/2}\exp\left\{- \frac{\alpha(x_{i} - x_{i-1} - h_n b_{i-1}(\mu))^2}{2a_{i-1}(\sigma)^2 h_n}\right\}, \\
    w_i^{(\gamma)} 
    & = \left(\frac{1 + \gamma}{2 \pi h_n a_{i -1}(\sigma)^2} \right)^{\gamma/2(1 + \gamma)}\exp\left\{- \frac{\gamma(x_{i} - x_{i-1} - h_n b_{i-1}(\mu))^2}{2a_{i-1}(\sigma)^2 h_n}\right\}. 
\end{align*}

\begin{figure}
    \begin{minipage}{0.48\textwidth}
     \centering
     \includegraphics[scale = 0.20]{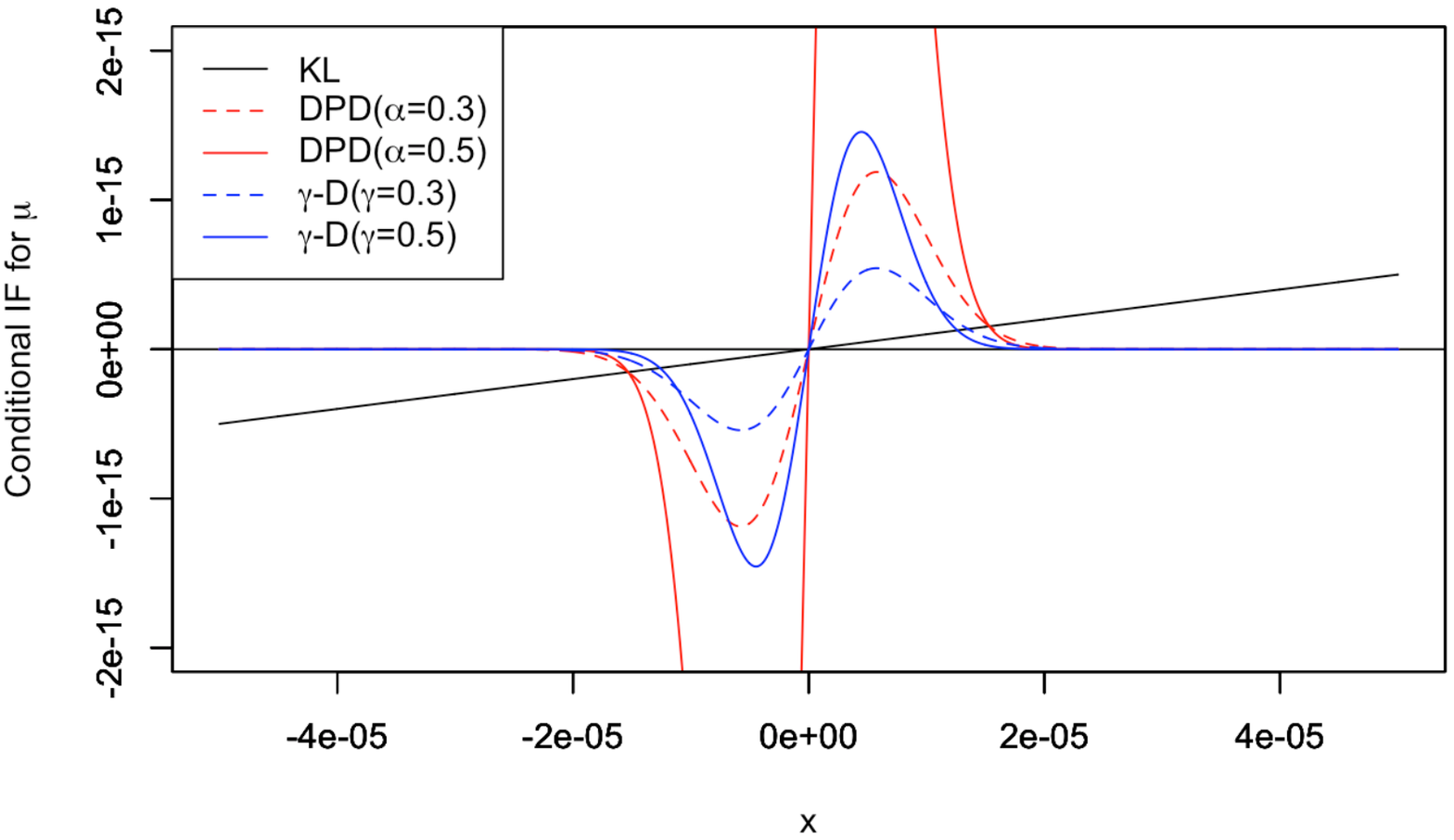}
     \end{minipage}
     \begin{minipage}{0.48\textwidth}
     \centering
     \includegraphics[scale = 0.20]{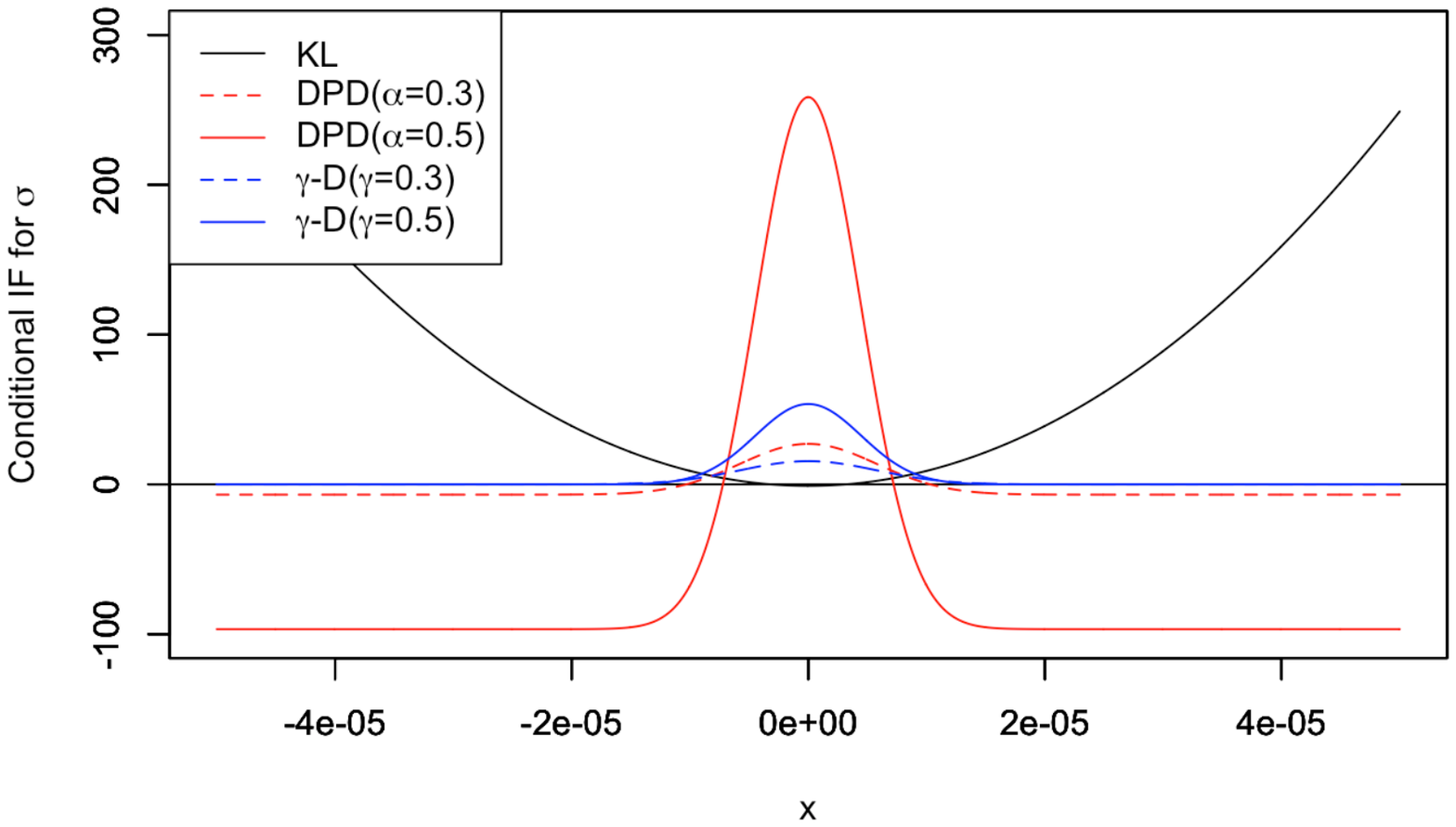}
     \end{minipage}
     \caption{The left figure is the conditional influence functions of $\mu$, and the right figure is that of $\sigma$ for the model A with $n = 100$, $h_n = n^{-0.55}$, $x_{i-1} = h_n$, $\mu_0 = 1.0$ and $\sigma_0 = 1.0$. The horizontal axis indicates $x_i$. }
     \label{fig:IFc_case1}
 \end{figure}

In Figure \ref{fig:IFc_case1}, the black line indicates the conditional influence function of MLE, the red line indicates the conditional influence function of the density power divergence-based estimator, and the blue line indicates the conditional influence function of the $\gamma$-divergence-based estimator. 
From Figure \ref{fig:IFc_case1}, it is easy to see that the conditional influence function of MLE is not bounded. 
On the other hand, the conditional influence function of the density power- and $\gamma$- divergence-based estimator seem to be bounded. 
In Figure \ref{fig:IFc_case1}, we note that the conditional influence functions of $\gamma$-divergence-based estimator seem to have the redescending properties as $x_i \to \infty$.

\section{Simulation studies}
\label{sec:sim}
In this section, we show the performance of the robust estimation using the divergences in discretely observed diffusion processes. Here, we set the discretisation step $h_n = n^{-0.55}$. 
We consider the following two diffusion models in each scenario. 
\begin{enumerate}[label = Model \Alph*:, ref = \Alph*, align = left]
\item \label{modelA} $a(x, \sigma) = \sigma, ~b(x, \mu) = -\mu x$. 
\item \label{modelB} $a(x, \sigma) =  1 + \sigma/(1 + x^{2}), ~ b(x, \mu) =  - \mu x$. 
\end{enumerate}
Note that the model \ref{modelA} is the Ornstein-Uhlenbeck process. 
We conduct two structures of generating outliers in each scenario. 
The configuration of the values of sample size $n$ and the parameter of divergence $\alpha$, $\gamma$ are $n = 50, 100, 200, 500$ and $\alpha, \gamma = 0.0, 0.3, 0.5$. 
For each of the configurations and the models, the bias and the mean square error (MSE) in each table are obtained by Monte Carlo simulation repeated $2000$ times. 
\subsection{Scenario 1}
\begin{figure}
     \centering
     \includegraphics[keepaspectratio, scale = 0.3]{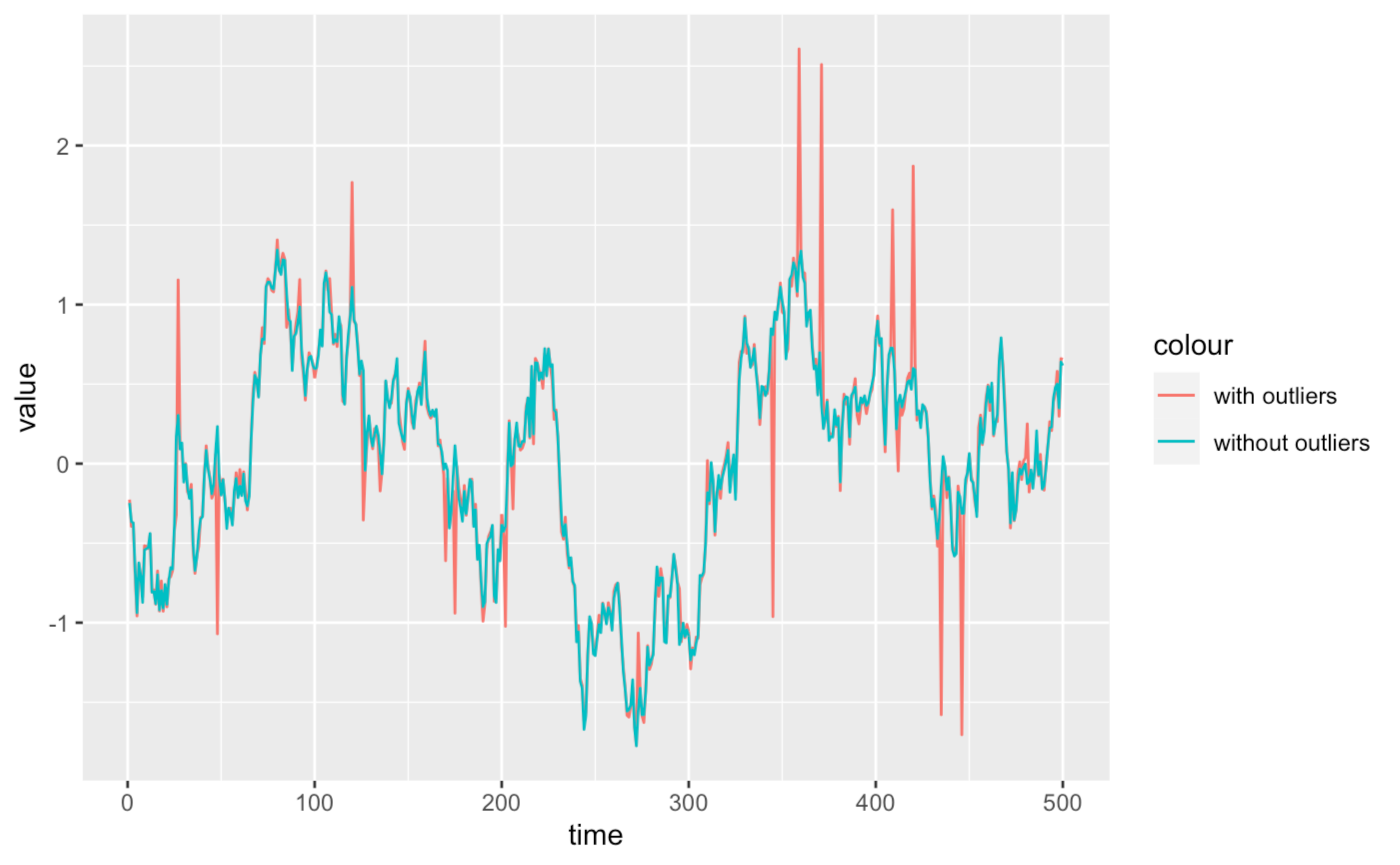}
     \caption{The blue and red lines are the sample path for the model \ref{modelA} with $\sigma_z^2 = 1.0$, $n = 500$, $\mu_0 = 1.0$ and $\sigma_0 = 1.0$. The blue line is the sample path without outliers, and the red line is the sample path without outliers. }
     \label{fig:data1}
 \end{figure}

Consider the AO structure as an outlier generator of time series. A special case of the AO model was originally introduced by \cite{fox1972outliers}. Let the observation $\{Y_{t_{i}^{n}}\}_{i=0,1,\ldots,n}$ be a wide-sense stationary "core" process of interest, and let $\{Z_{t_{i}^{n}}\}_{i=0,1,\ldots,n}$ be a stationary outlier process, and let $\{R_{t_{i}^{n}}\}_{i=0,1,\ldots,n}$ be a zero-one process with $\Pr(R_{t_i^n} = 0) = 1- \ep$ and $\Pr(R_{t_i^n} = 1) = \ep$. 
In practice, the fraction $\ep$ is often positive but small. 
Under an AO model, instead of $Y_{t_i^n}$ one actually observes
\begin{align*}
   Y_{t_i^n} =  X_{t_{i}^n} + R_{t_{i}^n}Z_{t_{i}^n},
\end{align*}
where the processes $\{X_{t_{i}^{n}}\}_{i=0,1,\ldots,n}$, $\{Z_{t_{i}^{n}}\}_{i=0,1,\ldots,n}$, and $\{R_{t_{i}^{n}}\}_{i=0,1,\ldots,n}$ are assumed to be independent of each other. 
Therefore, $Y_{t_{i}^{n}}$ is an outlier when $R_{t_i^n} = 1$.  
In this scenario, we assume that $Z_{t_{i}^n}$ is independent and identically distributed as normal distribution $N(0, \sigma_z^2)$ with $\sigma_z^2 = 1.0, 1.5$, and $\ep = 0.05$. 
The sample paths are given in Figure \ref{fig:data1}. 
Actually, the red sample path is observed and the blue one is not observed. 
We are interested in whether we can estimate the generating process of the blue sample path using the red one.

Tables \ref{tab:bias_sce1_case1} and \ref{tab:bias_sce1_case2} are given the biases for each model and the configurations. 
Moreover, Tables \ref{tab:MSE_sce1_case1} and \ref{tab:MSE_sce1_case2} are given the MSEs for each model and the configurations. 
From Tables \ref{tab:bias_sce1_case1}--\ref{tab:MSE_sce1_case2}, we can see that MLE is considerably influenced by outliers. 
On the other hand, the density power- and $\gamma$-divergences based estimators are little influenced by outliers. 
We also see that the density power- and $\gamma$-divergences based estimators and MLE have almost the same accuracy in the absence of outliers. 
Furthermore, from Tables \ref{tab:MSE_sce1_case1} and\ref{tab:MSE_sce1_case2}, it is seen that MSEs of the estimators in the absence of outliers decrease with $n$ increases, that is, the estimators are consistent in the absence of outliers. 
When outliers are included, MSE of MLE not only fails to decrease but actually increases as the sample size $n$ grows. 
Therefore, we see that MLE is not consistent. 
On the other hand, it is shown that the density power- and $\gamma$-divergences based estimators are consistent even when outliers are included from Theorem \ref{thm1}. 
Therefore, it is seen that MSEs of these estimators decrease with $n$ increases from Tables \ref{tab:MSE_sce1_case1} and \ref{tab:MSE_sce1_case2}.


\begin{table}[!tbp]
\caption{Empirical biases for the model \ref{modelA} in scenario 1 when $\mu_0 = 1.0$ and $\sigma_0 = 1.0$. \label{tab:bias_sce1_case1}} 
\begin{center}
\begin{tabular}{rrcrrcrrcrr}
\hline\hline
\multicolumn{2}{c}{\bfseries }&\multicolumn{1}{c}{\bfseries }&\multicolumn{2}{c}{\bfseries without outliers}&\multicolumn{1}{c}{\bfseries }&\multicolumn{2}{c}{\bfseries $\sigma_z^2 = 1.0$}&\multicolumn{1}{c}{\bfseries }&\multicolumn{2}{c}{\bfseries $\sigma_z^2 = 1.5$}\tabularnewline
\cline{4-5} \cline{7-8} \cline{10-11}
\multicolumn{1}{c}{$n$}&\multicolumn{1}{c}{$\alpha, \gamma$}&\multicolumn{1}{c}{}&\multicolumn{1}{c}{$\mu$}&\multicolumn{1}{c}{$\sigma$}&\multicolumn{1}{c}{}&\multicolumn{1}{c}{$\mu$}&\multicolumn{1}{c}{$\sigma$}&\multicolumn{1}{c}{}&\multicolumn{1}{c}{$\mu$}&\multicolumn{1}{c}{$\sigma$}\tabularnewline
\hline
&&&\multicolumn{8}{c}{\underline{\textbf{MLE}}}\tabularnewline
$ 50$&$0.0$&&$-0.0720$&$0.2383$&&$0.2222$&$1.0408$&&$0.3227$&$1.3100$\tabularnewline
$100$&$0.0$&&$-0.0452$&$0.1806$&&$0.3889$&$1.4168$&&$0.5380$&$1.8764$\tabularnewline
$200$&$0.0$&&$-0.0300$&$0.1322$&&$0.5991$&$1.9954$&&$0.8104$&$2.7239$\tabularnewline
$500$&$0.0$&&$-0.0174$&$0.1063$&&$0.9268$&$3.1066$&&$1.2292$&$4.3395$\tabularnewline
&&&\multicolumn{8}{c}{\underline{\textbf{Density power divergence based estimator}}}\tabularnewline
$ 50$&$0.3$&&$-0.0704$&$0.2407$&&$0.0475$&$0.6322$&&$0.0545$&$0.6750$\tabularnewline
$100$&$0.3$&&$-0.0446$&$0.1807$&&$0.0792$&$0.5497$&&$0.0788$&$0.5546$\tabularnewline
$200$&$0.3$&&$-0.0299$&$0.1246$&&$0.0922$&$0.4537$&&$0.0858$&$0.4318$\tabularnewline
$500$&$0.3$&&$-0.0170$&$0.1057$&&$0.0957$&$0.3662$&&$0.0865$&$0.3366$\tabularnewline
$ 50$&$0.5$&&$-0.0727$&$0.2417$&&$0.0162$&$0.5161$&&$0.0180$&$0.5208$\tabularnewline
$100$&$0.5$&&$-0.0460$&$0.1792$&&$0.0466$&$0.4287$&&$0.0453$&$0.4195$\tabularnewline
$200$&$0.5$&&$-0.0306$&$0.1200$&&$0.0603$&$0.3327$&&$0.0569$&$0.3149$\tabularnewline
$500$&$0.5$&&$-0.0171$&$0.1055$&&$0.0705$&$0.2725$&&$0.0661$&$0.2537$\tabularnewline
&&&\multicolumn{8}{c}{\underline{\textbf{$\gamma$-divergence based estimator}}}\tabularnewline
$ 50$&$0.3$&&$-0.0704$&$0.2406$&&$0.0432$&$0.6289$&&$0.0481$&$0.6682$\tabularnewline
$100$&$0.3$&&$-0.0446$&$0.1809$&&$0.0729$&$0.5448$&&$0.0702$&$0.5485$\tabularnewline
$200$&$0.3$&&$-0.0299$&$0.1245$&&$0.0834$&$0.4486$&&$0.0748$&$0.4252$\tabularnewline
$500$&$0.3$&&$-0.0170$&$0.1055$&&$0.0845$&$0.3613$&&$0.0734$&$0.3315$\tabularnewline
$ 50$&$0.5$&&$-0.0734$&$0.2418$&&$0.0047$&$0.5100$&&$0.0028$&$0.5115$\tabularnewline
$100$&$0.5$&&$-0.0464$&$0.1791$&&$0.0323$&$0.4234$&&$0.0274$&$0.4122$\tabularnewline
$200$&$0.5$&&$-0.0308$&$0.1195$&&$0.0423$&$0.3272$&&$0.0355$&$0.3079$\tabularnewline
$500$&$0.5$&&$-0.0172$&$0.1054$&&$0.0490$&$0.2675$&&$0.0418$&$0.2489$\tabularnewline
\hline
\end{tabular}\end{center}
\end{table}

\begin{table}[!tbp]
\caption{Empirical MSEs for the model \ref{modelA} in scenario 1  when $\mu_0 = 1.0$ and $\sigma_0 = 1.0$. \label{tab:MSE_sce1_case1}} 
\begin{center}
\begin{tabular}{rrcrrcrrcrr}
\hline\hline
\multicolumn{2}{c}{\bfseries }&\multicolumn{1}{c}{\bfseries }&\multicolumn{2}{c}{\bfseries without outliers}&\multicolumn{1}{c}{\bfseries }&\multicolumn{2}{c}{\bfseries $\sigma_z^2 = 1.0$}&\multicolumn{1}{c}{\bfseries }&\multicolumn{2}{c}{\bfseries $\sigma_z^2 = 1.5$}\tabularnewline
\cline{4-5} \cline{7-8} \cline{10-11}
\multicolumn{1}{c}{$n$}&\multicolumn{1}{c}{$\alpha, \gamma$}&\multicolumn{1}{c}{}&\multicolumn{1}{c}{$\mu$}&\multicolumn{1}{c}{$\sigma$}&\multicolumn{1}{c}{}&\multicolumn{1}{c}{$\mu$}&\multicolumn{1}{c}{$\sigma$}&\multicolumn{1}{c}{}&\multicolumn{1}{c}{$\mu$}&\multicolumn{1}{c}{$\sigma$}\tabularnewline
\hline
&&&\multicolumn{8}{c}{\underline{\textbf{MLE}}}\tabularnewline
$ 50$&$0.0$&&$0.0146$&$0.5570$&&$0.0639$&$1.6402$&&$0.1187$&$ 2.2732$\tabularnewline
$100$&$0.0$&&$0.0066$&$0.3880$&&$0.1579$&$2.3954$&&$0.2961$&$ 3.9091$\tabularnewline
$200$&$0.0$&&$0.0034$&$0.2529$&&$0.3624$&$4.2347$&&$0.6601$&$ 7.6726$\tabularnewline
$500$&$0.0$&&$0.0012$&$0.1624$&&$0.8602$&$9.8134$&&$1.5121$&$18.9940$\tabularnewline
&&&\multicolumn{8}{c}{\underline{\textbf{Density power divergence based estimator}}}\tabularnewline
$ 50$&$0.3$&&$0.0161$&$0.6051$&&$0.0267$&$1.4553$&&$0.0308$&$ 1.6694$\tabularnewline
$100$&$0.3$&&$0.0074$&$0.4135$&&$0.0192$&$1.0768$&&$0.0199$&$ 1.1419$\tabularnewline
$200$&$0.3$&&$0.0038$&$0.2659$&&$0.0156$&$0.6937$&&$0.0143$&$ 0.6610$\tabularnewline
$500$&$0.3$&&$0.0014$&$0.1682$&&$0.0118$&$0.3928$&&$0.0100$&$ 0.3620$\tabularnewline
$ 50$&$0.5$&&$0.0186$&$0.6568$&&$0.0221$&$1.2373$&&$0.0231$&$ 1.3035$\tabularnewline
$100$&$0.5$&&$0.0085$&$0.4437$&&$0.0131$&$0.8766$&&$0.0130$&$ 0.8772$\tabularnewline
$200$&$0.5$&&$0.0043$&$0.2860$&&$0.0094$&$0.5224$&&$0.0088$&$ 0.4996$\tabularnewline
$500$&$0.5$&&$0.0016$&$0.1800$&&$0.0072$&$0.3097$&&$0.0065$&$ 0.2939$\tabularnewline
&&&\multicolumn{8}{c}{\underline{\textbf{$\gamma$-divergence based estimator}}}\tabularnewline
$ 50$&$0.3$&&$0.0162$&$0.6051$&&$0.0261$&$1.4481$&&$0.0297$&$ 1.6495$\tabularnewline
$100$&$0.3$&&$0.0074$&$0.4137$&&$0.0180$&$1.0674$&&$0.0183$&$ 1.1287$\tabularnewline
$200$&$0.3$&&$0.0038$&$0.2659$&&$0.0138$&$0.6841$&&$0.0123$&$ 0.6490$\tabularnewline
$500$&$0.3$&&$0.0014$&$0.1681$&&$0.0097$&$0.3874$&&$0.0078$&$ 0.3569$\tabularnewline
$ 50$&$0.5$&&$0.0192$&$0.6578$&&$0.0223$&$1.2307$&&$0.0231$&$ 1.2845$\tabularnewline
$100$&$0.5$&&$0.0087$&$0.4439$&&$0.0120$&$0.8688$&&$0.0116$&$ 0.8657$\tabularnewline
$200$&$0.5$&&$0.0044$&$0.2853$&&$0.0074$&$0.5166$&&$0.0067$&$ 0.4924$\tabularnewline
$500$&$0.5$&&$0.0016$&$0.1800$&&$0.0046$&$0.3068$&&$0.0038$&$ 0.2915$\tabularnewline
\hline
\end{tabular}\end{center}
\end{table}

\begin{table}[!tbp]
\caption{Empirical biases for the model \ref{modelB} in scenario 1  when $\mu_0 = 1.0$ and $\sigma_0 = 1.0$. \label{tab:bias_sce1_case2}} 
\begin{center}
\begin{tabular}{rrcrrcrrcrr}
\hline\hline
\multicolumn{2}{c}{\bfseries }&\multicolumn{1}{c}{\bfseries }&\multicolumn{2}{c}{\bfseries without outliers}&\multicolumn{1}{c}{\bfseries }&\multicolumn{2}{c}{\bfseries $\sigma_z^2 = 1.0$}&\multicolumn{1}{c}{\bfseries }&\multicolumn{2}{c}{\bfseries $\sigma_z^2 = 1.5$}\tabularnewline
\cline{4-5} \cline{7-8} \cline{10-11}
\multicolumn{1}{c}{$n$}&\multicolumn{1}{c}{$\alpha, \gamma$}&\multicolumn{1}{c}{}&\multicolumn{1}{c}{$\mu$}&\multicolumn{1}{c}{$\sigma$}&\multicolumn{1}{c}{}&\multicolumn{1}{c}{$\mu$}&\multicolumn{1}{c}{$\sigma$}&\multicolumn{1}{c}{}&\multicolumn{1}{c}{$\mu$}&\multicolumn{1}{c}{$\sigma$}\tabularnewline
\hline
&&&\multicolumn{8}{c}{\underline{\textbf{MLE}}}\tabularnewline
$ 50$&$0.0$&&$-0.1672$&$ 0.1515$&&$0.2730$&$0.6262$&&$0.4734$&$0.8472$\tabularnewline
$100$&$0.0$&&$-0.1090$&$ 0.1171$&&$0.5286$&$0.8322$&&$0.8361$&$1.2111$\tabularnewline
$200$&$0.0$&&$-0.0730$&$ 0.0843$&&$0.8792$&$1.1909$&&$1.3658$&$1.8420$\tabularnewline
$500$&$0.0$&&$-0.0423$&$ 0.0713$&&$1.4996$&$1.9372$&&$2.3352$&$3.1722$\tabularnewline
&&&\multicolumn{8}{c}{\underline{\textbf{Density power divergence based estimator}}}\tabularnewline
$ 50$&$0.3$&&$-0.1487$&$ 0.0475$&&$0.0690$&$0.2639$&&$0.0965$&$0.2907$\tabularnewline
$100$&$0.3$&&$-0.1010$&$ 0.0046$&&$0.1587$&$0.2304$&&$0.1765$&$0.2398$\tabularnewline
$200$&$0.3$&&$-0.0705$&$-0.0356$&&$0.2158$&$0.1833$&&$0.2238$&$0.1802$\tabularnewline
$500$&$0.3$&&$-0.0414$&$-0.0426$&&$0.2623$&$0.1557$&&$0.2566$&$0.1425$\tabularnewline
$ 50$&$0.5$&&$-0.1444$&$ 0.0023$&&$0.0389$&$0.1699$&&$0.0543$&$0.1825$\tabularnewline
$100$&$0.5$&&$-0.0994$&$-0.0440$&&$0.1159$&$0.1286$&&$0.1240$&$0.1303$\tabularnewline
$200$&$0.5$&&$-0.0693$&$-0.0854$&&$0.1604$&$0.0753$&&$0.1618$&$0.0688$\tabularnewline
$500$&$0.5$&&$-0.0403$&$-0.0891$&&$0.1998$&$0.0523$&&$0.1938$&$0.0410$\tabularnewline
&&&\multicolumn{8}{c}{\underline{\textbf{$\gamma$-divergence based estimator}}}\tabularnewline
$ 50$&$0.3$&&$-0.1483$&$ 0.0459$&&$0.0640$&$0.2593$&&$0.0889$&$0.2850$\tabularnewline
$100$&$0.3$&&$-0.1007$&$ 0.0029$&&$0.1516$&$0.2261$&&$0.1662$&$0.2349$\tabularnewline
$200$&$0.3$&&$-0.0704$&$-0.0372$&&$0.2051$&$0.1790$&&$0.2093$&$0.1753$\tabularnewline
$500$&$0.3$&&$-0.0413$&$-0.0438$&&$0.2468$&$0.1515$&&$0.2368$&$0.1378$\tabularnewline
$ 50$&$0.5$&&$-0.1433$&$-0.0025$&&$0.0267$&$0.1601$&&$0.0369$&$0.1707$\tabularnewline
$100$&$0.5$&&$-0.0987$&$-0.0486$&&$0.0990$&$0.1190$&&$0.1009$&$0.1197$\tabularnewline
$200$&$0.5$&&$-0.0689$&$-0.0899$&&$0.1359$&$0.0655$&&$0.1301$&$0.0581$\tabularnewline
$500$&$0.5$&&$-0.0399$&$-0.0930$&&$0.1659$&$0.0425$&&$0.1528$&$0.0305$\tabularnewline
\hline
\end{tabular}\end{center}
\end{table}


\begin{table}[!tbp]
\caption{Empirical MSEs for the model \ref{modelB} in scenario 1  when $\mu_0 = 1.0$ and $\sigma_0 = 1.0$. \label{tab:MSE_sce1_case2}} 
\begin{center}
\begin{tabular}{rrcrrcrrcrr}
\hline\hline
\multicolumn{2}{c}{\bfseries }&\multicolumn{1}{c}{\bfseries }&\multicolumn{2}{c}{\bfseries without outliers}&\multicolumn{1}{c}{\bfseries }&\multicolumn{2}{c}{\bfseries $\sigma_z^2 = 1.0$}&\multicolumn{1}{c}{\bfseries }&\multicolumn{2}{c}{\bfseries $\sigma_z^2 = 1.5$}\tabularnewline
\cline{4-5} \cline{7-8} \cline{10-11}
\multicolumn{1}{c}{$n$}&\multicolumn{1}{c}{$\alpha, \gamma$}&\multicolumn{1}{c}{}&\multicolumn{1}{c}{$\mu$}&\multicolumn{1}{c}{$\sigma$}&\multicolumn{1}{c}{}&\multicolumn{1}{c}{$\mu$}&\multicolumn{1}{c}{$\sigma$}&\multicolumn{1}{c}{}&\multicolumn{1}{c}{$\mu$}&\multicolumn{1}{c}{$\sigma$}\tabularnewline
\hline
&&&\multicolumn{8}{c}{\underline{\textbf{MLE}}}\tabularnewline
$ 50$&$0.0$&&$0.0976$&$0.3761$&&$0.1721$&$0.7683$&&$0.3217$&$ 1.0938$\tabularnewline
$100$&$0.0$&&$0.0466$&$0.2586$&&$0.3260$&$0.9512$&&$0.7456$&$ 1.7253$\tabularnewline
$200$&$0.0$&&$0.0240$&$0.1643$&&$0.7970$&$1.5827$&&$1.8894$&$ 3.5575$\tabularnewline
$500$&$0.0$&&$0.0090$&$0.1052$&&$2.2577$&$3.8579$&&$5.4623$&$10.1679$\tabularnewline
&&&\multicolumn{8}{c}{\underline{\textbf{Density power divergence based estimator}}}\tabularnewline
$ 50$&$0.3$&&$0.1014$&$0.3438$&&$0.1350$&$0.5846$&&$0.1573$&$ 0.6600$\tabularnewline
$100$&$0.3$&&$0.0493$&$0.2265$&&$0.0971$&$0.4319$&&$0.1094$&$ 0.4559$\tabularnewline
$200$&$0.3$&&$0.0259$&$0.1420$&&$0.0891$&$0.2657$&&$0.0955$&$ 0.2647$\tabularnewline
$500$&$0.3$&&$0.0099$&$0.0888$&&$0.0877$&$0.1613$&&$0.0854$&$ 0.1552$\tabularnewline
$ 50$&$0.5$&&$0.1138$&$0.3554$&&$0.1366$&$0.5400$&&$0.1474$&$ 0.5814$\tabularnewline
$100$&$0.5$&&$0.0560$&$0.2343$&&$0.0857$&$0.3736$&&$0.0907$&$ 0.3889$\tabularnewline
$200$&$0.5$&&$0.0292$&$0.1515$&&$0.0661$&$0.2160$&&$0.0677$&$ 0.2135$\tabularnewline
$500$&$0.5$&&$0.0112$&$0.0963$&&$0.0569$&$0.1283$&&$0.0546$&$ 0.1256$\tabularnewline
&&&\multicolumn{8}{c}{\underline{\textbf{$\gamma$-divergence based estimator}}}\tabularnewline
$ 50$&$0.3$&&$0.1016$&$0.3435$&&$0.1329$&$0.5789$&&$0.1530$&$ 0.6507$\tabularnewline
$100$&$0.3$&&$0.0495$&$0.2263$&&$0.0940$&$0.4273$&&$0.1044$&$ 0.4498$\tabularnewline
$200$&$0.3$&&$0.0260$&$0.1420$&&$0.0838$&$0.2624$&&$0.0881$&$ 0.2612$\tabularnewline
$500$&$0.3$&&$0.0099$&$0.0888$&&$0.0793$&$0.1592$&&$0.0749$&$ 0.1530$\tabularnewline
$ 50$&$0.5$&&$0.1167$&$0.3550$&&$0.1372$&$0.5332$&&$0.1460$&$ 0.5693$\tabularnewline
$100$&$0.5$&&$0.0574$&$0.2341$&&$0.0823$&$0.3675$&&$0.0853$&$ 0.3815$\tabularnewline
$200$&$0.5$&&$0.0298$&$0.1518$&&$0.0584$&$0.2123$&&$0.0577$&$ 0.2098$\tabularnewline
$500$&$0.5$&&$0.0115$&$0.0967$&&$0.0441$&$0.1262$&&$0.0398$&$ 0.1237$\tabularnewline
\hline
\end{tabular}\end{center}
\end{table}

\subsection{Scenario 2}
Consider the RO structure as an outlier generator of time series. RO models have the form
\begin{align*}
    Y_{t_{i}^{n}}= (1 - R_{t_i^n})X_{t_{i}^{n}}+  R_{t_i^n} Z_{t_{i}^{n}},
\end{align*}
where $R_{t_i^n}$ is a zero-one process with $\Pr(R_{t_i^n} = 0) = 1- \ep$ and $\Pr(R_{t_i^n} = 1) = \ep$, and $Z_{t_i^n}$ is a "replacement" process that is not necessarily independent of $X_{t_i^n}$. 
Similar to scenario 1, we assume that $Z_{t_{i}^n}$ is independent and identically distributed as normal distribution $N(0, \sigma_z^2)$ with $\sigma_z^2 = 1.0, 1.5$, and $\ep = 0.05$. 

From Tables \ref{tab:bias_sce2_case1}--\ref{tab:MSE_sce2_case2}, similarly to scenario 1, we can see that MLE is considerably influenced by outliers. 
On the other hand, the density power- and $\gamma$-divergences based estimators are little influenced by outliers. 
We also see that the density power- and $\gamma$-divergences based estimators and MLE have almost the same accuracy in the absence of outliers. 
 

\begin{table}[!tbp]
\caption{Empirical biases for the model \ref{modelA} in scenario 2  when $\mu_0 = 1.0$ and $\sigma_0 = 1.0$. \label{tab:bias_sce2_case1}} 
\begin{center}
\begin{tabular}{rrcrrcrrcrr}
\hline\hline
\multicolumn{2}{c}{\bfseries }&\multicolumn{1}{c}{\bfseries }&\multicolumn{2}{c}{\bfseries without outliers}&\multicolumn{1}{c}{\bfseries }&\multicolumn{2}{c}{\bfseries $\sigma_z^2 = 1.0$}&\multicolumn{1}{c}{\bfseries }&\multicolumn{2}{c}{\bfseries $\sigma_z^2 = 1.5$}\tabularnewline
\cline{4-5} \cline{7-8} \cline{10-11}
\multicolumn{1}{c}{$n$}&\multicolumn{1}{c}{$\alpha, \gamma$}&\multicolumn{1}{c}{}&\multicolumn{1}{c}{$\mu$}&\multicolumn{1}{c}{$\sigma$}&\multicolumn{1}{c}{}&\multicolumn{1}{c}{$\mu$}&\multicolumn{1}{c}{$\sigma$}&\multicolumn{1}{c}{}&\multicolumn{1}{c}{$\mu$}&\multicolumn{1}{c}{$\sigma$}\tabularnewline
\hline
&&&\multicolumn{8}{c}{\underline{\textbf{MLE}}}\tabularnewline
$ 50$&$0.0$&&$-0.0720$&$0.2383$&&$0.2823$&$1.3795$&&$0.3726$&$1.6356$\tabularnewline
$100$&$0.0$&&$-0.0452$&$0.1806$&&$0.5010$&$1.9756$&&$0.6337$&$2.4232$\tabularnewline
$200$&$0.0$&&$-0.0300$&$0.1322$&&$0.7577$&$2.8292$&&$0.9417$&$3.5368$\tabularnewline
$500$&$0.0$&&$-0.0174$&$0.1063$&&$1.1807$&$4.5580$&&$1.4397$&$5.7696$\tabularnewline
&&&\multicolumn{8}{c}{\underline{\textbf{Density power divergence based estimator}}}\tabularnewline
$ 50$&$0.3$&&$-0.0704$&$0.2407$&&$0.0561$&$0.6537$&&$0.0594$&$0.6806$\tabularnewline
$100$&$0.3$&&$-0.0446$&$0.1807$&&$0.0798$&$0.5390$&&$0.0767$&$0.5332$\tabularnewline
$200$&$0.3$&&$-0.0299$&$0.1246$&&$0.0870$&$0.4211$&&$0.0808$&$0.4012$\tabularnewline
$500$&$0.3$&&$-0.0170$&$0.1057$&&$0.0864$&$0.3331$&&$0.0798$&$0.3157$\tabularnewline
$ 50$&$0.5$&&$-0.0727$&$0.2417$&&$0.0211$&$0.5221$&&$0.0211$&$0.5262$\tabularnewline
$100$&$0.5$&&$-0.0460$&$0.1792$&&$0.0465$&$0.4105$&&$0.0437$&$0.3991$\tabularnewline
$200$&$0.5$&&$-0.0306$&$0.1200$&&$0.0575$&$0.3088$&&$0.0547$&$0.2963$\tabularnewline
$500$&$0.5$&&$-0.0171$&$0.1055$&&$0.0659$&$0.2523$&&$0.0630$&$0.2437$\tabularnewline
&&&\multicolumn{8}{c}{\underline{\textbf{$\gamma$-divergence based estimator}}}\tabularnewline
$ 50$&$0.3$&&$-0.0704$&$0.2406$&&$0.0503$&$0.6489$&&$0.0522$&$0.6749$\tabularnewline
$100$&$0.3$&&$-0.0446$&$0.1809$&&$0.0716$&$0.5339$&&$0.0667$&$0.5265$\tabularnewline
$200$&$0.3$&&$-0.0299$&$0.1245$&&$0.0763$&$0.4144$&&$0.0686$&$0.3955$\tabularnewline
$500$&$0.3$&&$-0.0170$&$0.1055$&&$0.0735$&$0.3279$&&$0.0657$&$0.3110$\tabularnewline
$ 50$&$0.5$&&$-0.0734$&$0.2418$&&$0.0074$&$0.5151$&&$0.0042$&$0.5171$\tabularnewline
$100$&$0.5$&&$-0.0464$&$0.1791$&&$0.0292$&$0.4043$&&$0.0237$&$0.3918$\tabularnewline
$200$&$0.5$&&$-0.0308$&$0.1195$&&$0.0367$&$0.3027$&&$0.0316$&$0.2893$\tabularnewline
$500$&$0.5$&&$-0.0172$&$0.1054$&&$0.0419$&$0.2477$&&$0.0372$&$0.2393$\tabularnewline
\hline
\end{tabular}\end{center}
\end{table}


\begin{table}[!tbp]
\caption{Empirical MSEs for the model \ref{modelA} in scenario 2  when $\mu_0 = 1.0$ and $\sigma_0 = 1.0$. \label{tab:MSE_sce2_case1}} 
\begin{center}
\begin{tabular}{rrcrrcrrcrr}
\hline\hline
\multicolumn{2}{c}{\bfseries }&\multicolumn{1}{c}{\bfseries }&\multicolumn{2}{c}{\bfseries without outliers}&\multicolumn{1}{c}{\bfseries }&\multicolumn{2}{c}{\bfseries $\sigma_z^2 = 1.0$}&\multicolumn{1}{c}{\bfseries }&\multicolumn{2}{c}{\bfseries $\sigma_z^2 = 1.5$}\tabularnewline
\cline{4-5} \cline{7-8} \cline{10-11}
\multicolumn{1}{c}{$n$}&\multicolumn{1}{c}{$\alpha, \gamma$}&\multicolumn{1}{c}{}&\multicolumn{1}{c}{$\mu$}&\multicolumn{1}{c}{$\sigma$}&\multicolumn{1}{c}{}&\multicolumn{1}{c}{$\mu$}&\multicolumn{1}{c}{$\sigma$}&\multicolumn{1}{c}{}&\multicolumn{1}{c}{$\mu$}&\multicolumn{1}{c}{$\sigma$}\tabularnewline
\hline
&&&\multicolumn{8}{c}{\underline{\textbf{MLE}}}\tabularnewline
$ 50$&$0.0$&&$0.0146$&$0.5570$&&$0.0943$&$ 2.4601$&&$0.1534$&$ 3.2323$\tabularnewline
$100$&$0.0$&&$0.0066$&$0.3880$&&$0.2576$&$ 4.2911$&&$0.4082$&$ 6.2597$\tabularnewline
$200$&$0.0$&&$0.0034$&$0.2529$&&$0.5775$&$ 8.2573$&&$0.8903$&$12.7616$\tabularnewline
$500$&$0.0$&&$0.0012$&$0.1624$&&$1.3953$&$20.9381$&&$2.0740$&$33.4507$\tabularnewline
&&&\multicolumn{8}{c}{\underline{\textbf{Density power divergence based estimator}}}\tabularnewline
$ 50$&$0.3$&&$0.0161$&$0.6051$&&$0.0312$&$ 1.5489$&&$0.0338$&$ 1.7288$\tabularnewline
$100$&$0.3$&&$0.0074$&$0.4135$&&$0.0202$&$ 1.0342$&&$0.0200$&$ 1.0718$\tabularnewline
$200$&$0.3$&&$0.0038$&$0.2659$&&$0.0142$&$ 0.6244$&&$0.0130$&$ 0.5973$\tabularnewline
$500$&$0.3$&&$0.0014$&$0.1682$&&$0.0098$&$ 0.3570$&&$0.0086$&$ 0.3386$\tabularnewline
$ 50$&$0.5$&&$0.0186$&$0.6568$&&$0.0233$&$ 1.2671$&&$0.0235$&$ 1.3408$\tabularnewline
$100$&$0.5$&&$0.0085$&$0.4437$&&$0.0134$&$ 0.8215$&&$0.0130$&$ 0.8216$\tabularnewline
$200$&$0.5$&&$0.0043$&$0.2860$&&$0.0087$&$ 0.4861$&&$0.0084$&$ 0.4737$\tabularnewline
$500$&$0.5$&&$0.0016$&$0.1800$&&$0.0064$&$ 0.2915$&&$0.0060$&$ 0.2835$\tabularnewline
&&&\multicolumn{8}{c}{\underline{\textbf{$\gamma$-divergence based estimator}}}\tabularnewline
$ 50$&$0.3$&&$0.0162$&$0.6051$&&$0.0300$&$ 1.5393$&&$0.0325$&$ 1.7164$\tabularnewline
$100$&$0.3$&&$0.0074$&$0.4137$&&$0.0186$&$ 1.0250$&&$0.0181$&$ 1.0584$\tabularnewline
$200$&$0.3$&&$0.0038$&$0.2659$&&$0.0123$&$ 0.6137$&&$0.0109$&$ 0.5886$\tabularnewline
$500$&$0.3$&&$0.0014$&$0.1681$&&$0.0076$&$ 0.3519$&&$0.0065$&$ 0.3345$\tabularnewline
$ 50$&$0.5$&&$0.0192$&$0.6578$&&$0.0229$&$ 1.2566$&&$0.0231$&$ 1.3234$\tabularnewline
$100$&$0.5$&&$0.0087$&$0.4439$&&$0.0120$&$ 0.8149$&&$0.0114$&$ 0.8122$\tabularnewline
$200$&$0.5$&&$0.0044$&$0.2853$&&$0.0066$&$ 0.4810$&&$0.0062$&$ 0.4671$\tabularnewline
$500$&$0.5$&&$0.0016$&$0.1800$&&$0.0037$&$ 0.2894$&&$0.0033$&$ 0.2818$\tabularnewline
\hline
\end{tabular}\end{center}
\end{table}

\begin{table}[!tbp]
\caption{Empirical biases for the model \ref{modelB} in scenario 2  when $\mu_0 = 1.0$ and $\sigma_0 = 1.0$. \label{tab:bias_sce2_case2}} 
\begin{center}
\begin{tabular}{rrcrrcrrcrr}
\hline\hline
\multicolumn{2}{c}{\bfseries }&\multicolumn{1}{c}{\bfseries }&\multicolumn{2}{c}{\bfseries without outliers}&\multicolumn{1}{c}{\bfseries }&\multicolumn{2}{c}{\bfseries $\sigma_z^2 = 1.0$}&\multicolumn{1}{c}{\bfseries }&\multicolumn{2}{c}{\bfseries $\sigma_z^2 = 1.5$}\tabularnewline
\cline{4-5} \cline{7-8} \cline{10-11}
\multicolumn{1}{c}{$n$}&\multicolumn{1}{c}{$\alpha, \gamma$}&\multicolumn{1}{c}{}&\multicolumn{1}{c}{$\mu$}&\multicolumn{1}{c}{$\sigma$}&\multicolumn{1}{c}{}&\multicolumn{1}{c}{$\mu$}&\multicolumn{1}{c}{$\sigma$}&\multicolumn{1}{c}{}&\multicolumn{1}{c}{$\mu$}&\multicolumn{1}{c}{$\sigma$}\tabularnewline
\hline
&&&\multicolumn{8}{c}{\underline{\textbf{MLE}}}\tabularnewline
$ 50$&$0.0$&&$-0.1672$&$ 0.1515$&&$0.6031$&$0.9204$&&$0.8144$&$1.1395$\tabularnewline
$100$&$0.0$&&$-0.1090$&$ 0.1171$&&$1.1254$&$1.2922$&&$1.4698$&$1.6862$\tabularnewline
$200$&$0.0$&&$-0.0730$&$ 0.0843$&&$1.7963$&$1.8345$&&$2.3346$&$2.5017$\tabularnewline
$500$&$0.0$&&$-0.0423$&$ 0.0713$&&$3.0908$&$2.9636$&&$4.0263$&$4.2140$\tabularnewline
&&&\multicolumn{8}{c}{\underline{\textbf{Density power divergence based estimator}}}\tabularnewline
$ 50$&$0.3$&&$-0.1487$&$ 0.0475$&&$0.1346$&$0.3011$&&$0.1407$&$0.3075$\tabularnewline
$100$&$0.3$&&$-0.1010$&$ 0.0046$&&$0.2136$&$0.2403$&&$0.2113$&$0.2383$\tabularnewline
$200$&$0.3$&&$-0.0705$&$-0.0356$&&$0.2615$&$0.1805$&&$0.2488$&$0.1680$\tabularnewline
$500$&$0.3$&&$-0.0414$&$-0.0426$&&$0.2826$&$0.1395$&&$0.2624$&$0.1243$\tabularnewline
$ 50$&$0.5$&&$-0.1444$&$ 0.0023$&&$0.0837$&$0.1868$&&$0.0850$&$0.1857$\tabularnewline
$100$&$0.5$&&$-0.0994$&$-0.0440$&&$0.1501$&$0.1253$&&$0.1471$&$0.1219$\tabularnewline
$200$&$0.5$&&$-0.0693$&$-0.0854$&&$0.1865$&$0.0642$&&$0.1765$&$0.0549$\tabularnewline
$500$&$0.5$&&$-0.0403$&$-0.0891$&&$0.2095$&$0.0372$&&$0.1971$&$0.0268$\tabularnewline
&&&\multicolumn{8}{c}{\underline{\textbf{$\gamma$-divergence based estimator}}}\tabularnewline
$ 50$&$0.3$&&$-0.1483$&$ 0.0459$&&$0.1259$&$0.2977$&&$0.1298$&$0.3029$\tabularnewline
$100$&$0.3$&&$-0.1007$&$ 0.0029$&&$0.2011$&$0.2371$&&$0.1964$&$0.2344$\tabularnewline
$200$&$0.3$&&$-0.0704$&$-0.0372$&&$0.2437$&$0.1768$&&$0.2286$&$0.1637$\tabularnewline
$500$&$0.3$&&$-0.0413$&$-0.0438$&&$0.2575$&$0.1352$&&$0.2354$&$0.1200$\tabularnewline
$ 50$&$0.5$&&$-0.1433$&$-0.0025$&&$0.0638$&$0.1783$&&$0.0612$&$0.1757$\tabularnewline
$100$&$0.5$&&$-0.0987$&$-0.0486$&&$0.1219$&$0.1170$&&$0.1148$&$0.1124$\tabularnewline
$200$&$0.5$&&$-0.0689$&$-0.0899$&&$0.1478$&$0.0548$&&$0.1339$&$0.0447$\tabularnewline
$500$&$0.5$&&$-0.0399$&$-0.0930$&&$0.1589$&$0.0275$&&$0.1437$&$0.0165$\tabularnewline
\hline
\end{tabular}\end{center}
\end{table}

\begin{table}[!tbp]
\caption{Empirical MSEs for the model \ref{modelB} in scenario 2  when $\mu_0 = 1.0$ and $\sigma_0 = 1.0$. \label{tab:MSE_sce2_case2}} 
\begin{center}
\begin{tabular}{rrcrrcrrcrr}
\hline\hline
\multicolumn{2}{c}{\bfseries }&\multicolumn{1}{c}{\bfseries }&\multicolumn{2}{c}{\bfseries without outliers}&\multicolumn{1}{c}{\bfseries }&\multicolumn{2}{c}{\bfseries $\sigma_z^2 = 1.0$}&\multicolumn{1}{c}{\bfseries }&\multicolumn{2}{c}{\bfseries $\sigma_z^2 = 1.5$}\tabularnewline
\cline{4-5} \cline{7-8} \cline{10-11}
\multicolumn{1}{c}{$n$}&\multicolumn{1}{c}{$\alpha, \gamma$}&\multicolumn{1}{c}{}&\multicolumn{1}{c}{$\mu$}&\multicolumn{1}{c}{$\sigma$}&\multicolumn{1}{c}{}&\multicolumn{1}{c}{$\mu$}&\multicolumn{1}{c}{$\sigma$}&\multicolumn{1}{c}{}&\multicolumn{1}{c}{$\mu$}&\multicolumn{1}{c}{$\sigma$}\tabularnewline
\hline
&&&\multicolumn{8}{c}{\underline{\textbf{MLE}}}\tabularnewline
$ 50$&$0.0$&&$0.0976$&$0.3761$&&$0.4614$&$1.2232$&&$ 0.7609$&$ 1.6746$\tabularnewline
$100$&$0.0$&&$0.0466$&$0.2586$&&$1.3131$&$1.9285$&&$ 2.2070$&$ 3.1019$\tabularnewline
$200$&$0.0$&&$0.0240$&$0.1643$&&$3.2508$&$3.5298$&&$ 5.4742$&$ 6.4226$\tabularnewline
$500$&$0.0$&&$0.0090$&$0.1052$&&$9.5617$&$8.8883$&&$16.2201$&$17.8629$\tabularnewline
&&&\multicolumn{8}{c}{\underline{\textbf{Density power divergence based estimator}}}\tabularnewline
$ 50$&$0.3$&&$0.1014$&$0.3438$&&$0.1841$&$0.6493$&&$ 0.1923$&$ 0.6924$\tabularnewline
$100$&$0.3$&&$0.0493$&$0.2265$&&$0.1343$&$0.4228$&&$ 0.1357$&$ 0.4358$\tabularnewline
$200$&$0.3$&&$0.0259$&$0.1420$&&$0.1193$&$0.2599$&&$ 0.1121$&$ 0.2528$\tabularnewline
$500$&$0.3$&&$0.0099$&$0.0888$&&$0.1004$&$0.1548$&&$ 0.0885$&$ 0.1454$\tabularnewline
$ 50$&$0.5$&&$0.1138$&$0.3554$&&$0.1613$&$0.5560$&&$ 0.1634$&$ 0.5682$\tabularnewline
$100$&$0.5$&&$0.0560$&$0.2343$&&$0.1040$&$0.3577$&&$ 0.1037$&$ 0.3641$\tabularnewline
$200$&$0.5$&&$0.0292$&$0.1515$&&$0.0784$&$0.2115$&&$ 0.0738$&$ 0.2068$\tabularnewline
$500$&$0.5$&&$0.0112$&$0.0963$&&$0.0611$&$0.1241$&&$ 0.0555$&$ 0.1198$\tabularnewline
&&&\multicolumn{8}{c}{\underline{\textbf{$\gamma$-divergence based estimator}}}\tabularnewline
$ 50$&$0.3$&&$0.1016$&$0.3435$&&$0.1785$&$0.6453$&&$ 0.1851$&$ 0.6846$\tabularnewline
$100$&$0.3$&&$0.0495$&$0.2263$&&$0.1272$&$0.4204$&&$ 0.1273$&$ 0.4322$\tabularnewline
$200$&$0.3$&&$0.0260$&$0.1420$&&$0.1088$&$0.2579$&&$ 0.1007$&$ 0.2503$\tabularnewline
$500$&$0.3$&&$0.0099$&$0.0888$&&$0.0859$&$0.1528$&&$ 0.0740$&$ 0.1439$\tabularnewline
$ 50$&$0.5$&&$0.1167$&$0.3550$&&$0.1575$&$0.5506$&&$ 0.1583$&$ 0.5597$\tabularnewline
$100$&$0.5$&&$0.0574$&$0.2341$&&$0.0956$&$0.3546$&&$ 0.0939$&$ 0.3598$\tabularnewline
$200$&$0.5$&&$0.0298$&$0.1518$&&$0.0641$&$0.2094$&&$ 0.0589$&$ 0.2046$\tabularnewline
$500$&$0.5$&&$0.0115$&$0.0967$&&$0.0417$&$0.1229$&&$ 0.0365$&$ 0.1187$\tabularnewline
\hline
\end{tabular}\end{center}
\end{table}

\section{Concluding remarks}

This paper presents a robust estimation of the diffusion process by using divergences. 
We proposed the $\gamma$-divergence based estimation for the diffusion process model similarly to \cite{lee2013minimum, song2007minimum, song2017robust}, and proved the consistency and asymptotic normality of the proposed estimators. 
We derived the conditional influence function for the robust divergence-based estimators, and see the boundness in numerical studies. 

\section*{Acknowledgments}
This work was JSPS Grant-in-Aid for Early-Career Scientists Grant Number JP23K13019 and JP23K16852.

\begin{appendices}
\def\thesection{A.\arabic{section}}
\setcounter{equation}{0}
\def\theequation{A.\arabic{equation}}
\def\thethm{A.\arabic{thm}}
\def\thelem{A.\arabic{lem}}

\section{Proof of Theorem \ref{thm1}}
\label{A.1}

We assume $\gamma>0$ because the proof of the case of $\gamma=0$ is similar to that of $\gamma>0$. First of all, we will prove that 
\begin{align}
\frac{Q_{n,\gamma}(\theta)}{n}\xrightarrow{p} -\sigma^{-\gamma/(1+\gamma)}\left(1+\frac{\gamma\sigma_{0}^{2}}{\sigma^{2}}\right)^{-1/2}
\label{consistency}
\end{align}
uniformly in $\theta$. From the Euler's approximation, we have
\begin{align*}
X_{t_{i}^{n}}=X_{t_{i-1}^{n}}+b_{i-1}(\mu_{0})h_{n}+\sigma_{0}Z_{n,i}\sqrt{h_{n}}+\Delta_{n,i},
\end{align*}
where 
\begin{align*}
Z_{n,i}=\frac{1}{\sqrt{h_{n}}}(W_{t_{i}^{n}}-W_{t_{i-1}^{n}}),\quad \Delta_{n,i}=\int_{t_{i-1}^{n}}^{t_{i}^{n}}\left(b_{s}(\mu_{0})-b_{i-1}(\mu_{0})\right)ds.
\end{align*}
Let us define
\begin{align*}
K_{n,i}(\mu,\sigma):=&\frac{\gamma(X_{t_{i}^{n}}-X_{t_{i-1}^{n}}-h_{n}b_{i-1}(\mu))^{2}}{2h_{n}\sigma^{2}}-\frac{\gamma\sigma_{0}^{2}Z_{n,i}^{2}}{2\sigma^{2}} \\
=&\frac{\gamma B_{i-1}(\mu)^{2}h_{n}}{2\sigma^{2}}+\frac{\gamma\Delta_{n,i}^{2}}{2h_{n}\sigma^{2}}+\frac{\gamma\sigma_{0}B_{i-1}(\mu)Z_{n,i}\sqrt{h_{n}}}{\sigma^{2}}+\frac{\gamma B_{i-1}(\mu)\Delta_{n,i}}{\sigma^{2}} \\
&+\frac{\gamma\sigma_{0}Z_{n,i}\Delta_{n,i}}{\sigma^{2}\sqrt{h_{n}}},
\end{align*}
where $B_{i-1}(\mu):=b_{i-1}(\mu_{0})-b_{i-1}(\mu)$. Similarly to the proof of Theorem 1 of \cite{lee2013minimum}, the standard estimation gives us 
\begin{align*}
\sup_{\theta}\max_{1\leq i \leq n}|K_{n,i}(\mu,\sigma)|=o_{p}(h_{n}^{r}),\quad 0\leq r<\frac{1}{2},
\end{align*}
and
\begin{align*}
&\sup_{\theta}\left|-\sigma^{-\gamma/(1+\gamma)}\frac{1}{n}\sum_{i=1}^{n}\left[{\rm exp}\left\{-\frac{\gamma\left(X_{t_{i}^{n}}-X_{t_{i-1}^{n}}-h_{n}b_{i-1}(\mu)\right)^{2}}{2h_{n}\sigma^{2}}\right\}-{\rm exp}\left(-\frac{\gamma\sigma_{0}^{2}}{2\sigma^{2}}Z_{n,i}^{2}\right)\right]\right| \\
&=\sup_{\theta}\left|-\sigma^{-\gamma/(1+\gamma)}\frac{1}{n}\sum_{i=1}^{n}{\rm exp}\left(-\frac{\gamma\sigma_{0}^{2}}{2\sigma^{2}}Z_{n,i}^{2}\right)\left(e^{-K_{n,i}(\mu,\sigma)}-1\right)\right|\\
&\leq \sup_{\theta}\sigma^{-\gamma/(1+\gamma)}\max_{1\leq i\leq n}\left|e^{-K_{n,i}(\mu,\sigma)}-1\right|=o_{p}(1).
\end{align*}
Hence form Lemma 4 in \cite{lee2013minimum}, we have that uniformly in $\theta$, 
\begin{align*}
\frac{Q_{n,\gamma}(\theta)}{n}&=-\sigma^{-\gamma/(1+\gamma)}\frac{1}{n}\sum_{i=1}^{n}{\rm exp}\left\{-\frac{\gamma\left(X_{t_{i}^{n}}-X_{t_{i-1}^{n}}-h_{n}b_{i-1}(\mu)\right)^{2}}{2h_{n}\sigma^{2}}\right\}\\
&\xrightarrow{p}-\sigma^{-\gamma/(1+\gamma)}\left(1+\frac{\gamma\sigma_{0}^{2}}{\sigma^{2}}\right)^{-1/2}
\end{align*}
which establish \eqref{consistency}. Note that 
\begin{align*}
Q_{\gamma}(\sigma):=-\sigma^{-\gamma/(1+\gamma)}\left(1+\frac{\gamma\sigma_{0}^{2}}{\sigma^{2}}\right)^{-1/2}
\end{align*}
has a minimal value at $\sigma=\sigma_{0}$.

Next, we will verify that
\begin{align*}
&\frac{Q_{n,\gamma}(\theta)}{nh_{n}}-\frac{Q_{n,\gamma}(\mu_{0},\sigma)}{nh_{n}}\\
\xrightarrow{p}&\frac{\gamma\sigma^{-\frac{2+3\gamma}{1+\gamma}}}{2}
\left(1+\frac{\gamma\sigma_{0}^{2}}{\sigma^{2}}\right)^{-3/2}\int\left(b(x,\mu)-b(x,\mu_{0})\right)^{2}d\nu_{0}(x).
\end{align*}
From the Taylor expansion, we have
\begin{align*}
\begin{split}
&e^{-K_{n,i}(\mu,\sigma)}-e^{-K_{n,i}(\mu_{0},\sigma)}\\
=&-\frac{\gamma B_{i-1}(\mu)^{2}h_{n}}{2\sigma^{2}}-\frac{\gamma\sigma_{0}B_{i-1}(\mu)Z_{n,i}\sqrt{h_{n}}}{\sigma^{2}}-\frac{\gamma B_{i-1}(\mu)\Delta_{n,i}}{\sigma^{2}}+\frac{1}{2}K_{n,i}(\mu,\sigma)^{2}\\
&+\frac{K_{n,i}(\mu_{0},\sigma)^{2}}{2!}e^{\xi_{1,i}}+\frac{K_{n,i}(\mu,\sigma)^{3}}{3!}e^{\xi_{2,i}} \\
=&-\frac{\gamma\sigma_{0}B_{i-1}(\mu)Z_{n,i}\sqrt{h_{n}}}{\sigma^{2}}-\frac{\gamma}{2\sigma^{2}}\left(1-\frac{\gamma\sigma_{0}^{2}}{\sigma^{2}}Z_{n,i}^{2}\right)B_{i-1}(\mu)^{2}h_{n}+H_{n,i}(\mu,\sigma) \\
&+\frac{K_{n,i}(\mu_{0},\sigma)^{2}}{2!}e^{\xi_{1,i}}+\frac{K_{n,i}(\mu,\sigma)^{3}}{3!}e^{\xi_{2,i}},
\end{split}
\end{align*}
where $|\xi_{1,i}|\leq |K_{n,i}(\mu_{0},\sigma)|$, $|\xi_{2,i}|\leq |K_{n,i}(\mu,\sigma)|$, and 
\begin{align*}
H_{n,i}(\mu,\sigma):=-\frac{\gamma B_{i-1}(\mu)\Delta_{n,i}}{\sigma^{2}}+\frac{1}{2}K_{n,i}(\mu,\sigma)^{2}-\frac{1}{2}\frac{\gamma^{2}\sigma_{0}^{2}}{\sigma^{4}}Z_{n,i}^{2}B_{i-1}(\mu)^{2}h_{n}.
\end{align*}
Let us define
\begin{align*}
R_{n,i}(\mu,\sigma):=H_{n,i}(\mu,\sigma)+\frac{K_{n,i}(\mu_{0},\sigma)^{2}}{2!}e^{\xi_{1,i}}+\frac{K_{n,i}(\mu,\sigma)^{3}}{3!}e^{\xi_{2,i}}.
\end{align*}
Then, from Lemmas 2, 4 and 5 in \cite{lee2013minimum},  we have
\begin{align*}
\sup_{\theta}\max_{1\leq i\leq n}|H_{n,i}(\mu,\sigma)|=&o_{p}(h_{n}^{r_{1}}),\quad 0\leq r_{1}<1.5,\\
\sup_{\theta}\max_{1\leq i\leq n}|K_{n,i}(\mu,\sigma)^{3}|=&o_{p}(h_{n}^{r_{2}}),\quad 0\leq r_{2}<1.5,\\
\sup_{\theta}\max_{1\leq i\leq n}|K_{n,i}(\mu_{0},\sigma)^{2}|=&o_{p}(h_{n}^{r_{3}}),\quad 0\leq r_{3}<2,\\
\sup_{\theta}\max_{1\leq i\leq n}|e^{\xi_{1,i}}|\leq&\ {\rm exp}\left\{\sup_{\theta}\max_{1\leq i\leq n}|K_{n,i}(\mu_{0},\sigma)|\right\}=O_{p}(1),\\
\sup_{\theta}\max_{1\leq i\leq n}|e^{\xi_{2,i}}|\leq&\ {\rm exp}\left\{\sup_{\theta}\max_{1\leq i\leq n}|K_{n,i}(\mu,\sigma)|\right\}=O_{p}(1),\\
\sup_{\theta}\max_{1\leq i\leq n}R_{n,i}(\mu,\sigma)=&o_{p}(h_{n}^{r}),\quad 0\leq r<1.5,
\end{align*}
and therefore
\begin{align*}
&\frac{Q_{n,\gamma}(\theta)}{nh_{n}}-\frac{Q_{n,\gamma}(\mu_{0},\sigma)}{nh_{n}}\\
=&-\frac{\sigma^{-\gamma/(1+\gamma)}}{nh_{n}}\sum_{i=1}^{n}{\rm exp}\left(-\frac{\gamma\sigma_{0}^{2}}{2\sigma^{2}}Z_{n,i}^{2}\right)\left(e^{-K_{n,i}(\mu,\sigma)}-e^{-K_{n,i}(\mu_{0},\sigma)}\right) \\
=&\frac{\gamma\sigma^{-\frac{2+3\gamma}{1+\gamma}}\sigma_{0}}{n\sqrt{h_{n}}}\sum_{i=1}^{n}B_{i-1}(\mu)Z_{n,i}{\rm exp}\left(-\frac{\gamma\sigma_{0}^{2}}{2\sigma^{2}}Z_{n,i}^{2}\right)\\
&+\frac{\gamma\sigma^{-\frac{2+3\gamma}{1+\gamma}}}{2n}\sum_{i=1}^{n}\left(1-\frac{\gamma\sigma_{0}^{2}}{\sigma^{2}}Z_{n,i}^{2}\right)B_{i-1}(\mu)^{2}{\rm exp}\left(-\frac{\gamma\sigma_{0}^{2}}{2\sigma^{2}}Z_{n,i}^{2}\right)\\
&-\frac{\sigma^{-\gamma/(1+\gamma)}}{nh_{n}}\sum_{i=1}^{n}R_{n,i}(\mu,\sigma){\rm exp}\left(-\frac{\gamma\sigma_{0}^{2}}{2\sigma^{2}}Z_{n,i}^{2}\right)\\
\xrightarrow{p}&\frac{\gamma\sigma^{-\frac{2+3\gamma}{1+\gamma}}}{2}
\left(1+\frac{\gamma\sigma_{0}^{2}}{\sigma^{2}}\right)^{-3/2}\int\left\{b(x,\mu)-b(x,\mu_{0})\right\}^{2}d\nu_{0}(x)
\end{align*}
uniformly in $\theta$. This completes the proof of $\hat{\theta}_{n}^{(\gamma)}\xrightarrow{p}\theta_{0}$.

\medskip

Second, we will derive the asymptotic distribution of $\left(\sqrt{nh_{n}}(\hat{\mu}_{n}^{(\gamma)}-\mu_{0}),\right.$\\
$\left.\sqrt{n}(\hat{\sigma}_{n}^{(\gamma)}-\sigma_{0})\right)$. From the Taylor expansion, we get
\begin{align*}
0=&\left(
\begin{array}{c}
\partial_{\mu}Q_{n,\gamma}(\theta_{0})\\
\partial_{\sigma}Q_{n,\gamma}(\theta_{0})
\end{array}
\right)  + \int_{0}^{1}
\nabla^{2}Q_{n,\gamma}\left(\theta_{0}+u(\hat{\theta}_{n}^{(\gamma)}-\theta_{0})\right)du
\left(
\begin{array}{c}
\hat{\mu}_{n}^{(\gamma)}-\mu_{0}\\
\hat{\sigma}_{n}^{(\gamma)}-\sigma_{0}
\end{array}
\right)
\end{align*}
from which we can write that
\begin{align*}
L_{n,\gamma}=\int_{0}^{1}C_{n,\gamma}\left(\theta_{0}+u(\hat{\theta}_{n}^{(\gamma)}-\theta_{0})\right)du
\left(
\begin{array}{c}
\sqrt{nh_{n}}(\hat{\mu}_{n}^{(\gamma)}-\mu_{0})\\
\sqrt{n}(\hat{\sigma}_{n}^{(\gamma)}-\sigma_{0})
\end{array}
\right),
\end{align*}
where
\begin{align*}
L_{n,\gamma}&=\left(
\begin{array}{c}
\displaystyle -\frac{1}{\sqrt{nh_{n}}}\partial_{\mu}Q_{n,\gamma}(\theta_{0})\\
\displaystyle -\frac{1}{\sqrt{n}}\partial_{\sigma}Q_{n,\gamma}(\theta_{0})
\end{array}
\right), \\
C_{n,\gamma}(\theta)&=\left(
\begin{array}{cc}
\displaystyle \frac{1}{nh_{n}}\partial_{\mu}^{2}Q_{n,\gamma}(\theta) & \displaystyle \frac{1}{n\sqrt{h_{n}}}\partial_{\mu\sigma}^{2}Q_{n,\gamma}(\theta)\\
\displaystyle \frac{1}{n\sqrt{h_{n}}}\partial_{\sigma\mu}^{2}Q_{n,\gamma}(\theta) & \displaystyle \frac{1}{n}\partial_{\sigma}^{2}Q_{n,\gamma}(\theta)
\end{array}
\right).
\end{align*}
Let us define
\begin{align*}
m_{n,i}^{(\gamma)}(\theta)=-\sigma^{-\gamma/(1+\gamma)}{\rm exp}\left\{-\frac{\gamma\left(B_{i-1}(\mu)h_{n}+\sigma_{0}Z_{n,i}\sqrt{h_{n}}\right)^{2}}{2h_{n}\sigma^{2}}\right\},
\end{align*}
and
\begin{align*}
\zeta_{n,i}=
\left(
\begin{array}{c}
\frac{1}{\sqrt{nh_{n}}}\partial_{\mu}m_{n,i}^{(\gamma)}(\theta_{0}) \\
\frac{1}{\sqrt{n}}\partial_{\sigma}m_{n,i}^{(\gamma)}(\theta_{0})
\end{array}
\right).
\end{align*}
We note that
\begin{align*}
\partial_{\mu}m_{m,i}^{(\gamma)}(\theta_{0})&=-\sqrt{h_{n}}\gamma\sigma_{0}^{-\frac{1+2\gamma}{1+\gamma}}Z_{n,i}\partial_{\mu}b_{i-1}(\mu_{0})e^{-\frac{\gamma}{2}Z_{n,i}^{2}}, \\
\partial_{\sigma}m_{m,i}^{(\gamma)}(\theta_{0})&=\gamma\sigma_{0}^{-\frac{1+2\gamma}{1+\gamma}}\left(\frac{1}{1+\gamma}-Z_{n,i}^{2}\right)e^{-\frac{\gamma}{2}Z_{n,i}^{2}}.
\end{align*}
Utilizing the arguments:
\begin{align*}
E_{\theta_{0}}\left[Z_{n,i}e^{-\frac{\gamma}{2}Z_{n,i}^{2}}|\mathscr{G}_{i-1}^{n}\right]&=0,\\
E_{\theta_{0}}\left[Z_{n,i}^{2}e^{-\gamma Z_{n,i}^{2}}|\mathscr{G}_{i-1}^{n}\right]&=(1+2\gamma)^{-3/2},\\
E_{\theta_{0}}\left[\left(\frac{1}{1+\gamma}-Z_{n,i}^{2}\right)e^{-\frac{\gamma}{2}Z_{n,i}^{2}}|\mathscr{G}_{i-1}^{n}\right]&=0, \\
E_{\theta_{0}}\left[Z_{n,i}\left(\frac{1}{1+\gamma}-Z_{n,i}^{2}\right)e^{-\gamma Z_{n,i}^{2}}|\mathscr{G}_{i-1}^{n}\right]&=0, \\
E_{\theta_{0}}\left[\left(\frac{1}{1+\gamma}-Z_{n,i}^{2}\right)^{2}e^{-\gamma Z_{n,i}^{2}}|\mathscr{G}_{i-1}^{n}\right]&=\frac{3\gamma^{2}+4\gamma+2}{(1+2\gamma)^{5/2}(1 + \gamma)^2},
\end{align*}
where $\mathscr{G}_{i}^{n}$ denotes the sigma field by $\{w_{s}:s\leq t_{i}^{n}\}$, and Lemma 6 in \cite{lee2013minimum}, we can show 
\begin{align*}
\sum_{i=1}^{n}E_{\theta_{0}}[\zeta_{n,i}|\mathscr{G}_{i-1}^{n}]\xrightarrow{p}0, \\
\sum_{i=1}^{n}E_{\theta_{0}}[\zeta_{n,i}\zeta_{n,i}^{T}|\mathscr{G}_{i-1}^{n}]\xrightarrow{p}\Sigma^{(\gamma)}_{1}, \\
\sum_{i=1}^{n}E_{\theta_{0}}[|\zeta_{n,i}|^{4}|\mathscr{G}_{i-1}^{n}]\xrightarrow{p}0,
\end{align*}
where
\begin{align*}
\Sigma^{(\gamma)}_{1}=\gamma^{2}\sigma_{0}^{-\frac{2(1+2\gamma)}{1+\gamma}}
\left(
\begin{array}{cc}
\frac{1}{(1+2\gamma)^{3/2}}S & 0 \\
0 & \frac{3\gamma^{2}+4\gamma+2}{(1+2\gamma)^{5/2}(1 + \gamma)^2}
\end{array}
\right),
\end{align*}
and $S$ is the one defined in the assumption \ref{A7}. This means that $L_{n,\gamma}\xrightarrow{\mathcal{L}}N_{p+1}(0,\Sigma^{(\gamma)}_{1})$.

\medskip

Now, the proof of theorem is completed if we have
\begin{align}
C_{n,\gamma}(\theta_{0})\xrightarrow{p}\sigma_{0}^{-\frac{2+3\gamma}{1+\gamma}}
\left(
\begin{array}{cc}
\gamma(1+\gamma)^{-3/2}S & 0 \\
0 & 2\gamma(1+\gamma)^{-5/2}
\end{array}
\right).
\label{eq:C_lim}
\end{align}
We note that
\begin{align*}
\partial^{2}_{\mu}Q_{n,\gamma}(\theta)=&-\gamma\sigma^{-\frac{2+3\gamma}{1+\gamma}}\sum_{i=1}^{n}\left\{\left(X_{t_{i}^{n}}-X_{t_{i-1}^{n}}-h_{n}b_{i-1}(\mu)\right)^{2}\partial_{\mu}b_{i-1}(\mu)\partial_{\mu^{\top}}b_{i-1}(\mu)\frac{\gamma}{\sigma^{2}}\right.\\
&\left.-h_{n}\partial_{\mu}b_{i-1}(\mu)\partial_{\mu^{\top}}b_{i-1}(\mu)+\left(X_{t_{i}^{n}}-X_{t_{i-1}^{n}}-h_{n}b_{i-1}(\mu)\right)\partial^{2}_{\mu}b_{i-1}(\mu)\right\}\\
&\times{\rm exp}\left(-\frac{\gamma\sigma^{2}_{0}Z_{n,i}^{2}}{2\sigma^{2}}\right){\rm exp}\left(-K_{n,i}(\mu,\sigma)\right) \\
=&-\gamma\sigma^{-\frac{2+3\gamma}{1+\gamma}}\sum_{i=1}^{n}\sigma_{0}\partial^{2}_{\mu}b_{i-1}(\mu)Z_{n,i}\sqrt{h_{n}}{\rm exp}\left(-\frac{\gamma\sigma^{2}_{0}Z_{n,i}^{2}}{2\sigma^{2}}\right)\\
&-\gamma\sigma^{-\frac{2+3\gamma}{1+\gamma}}\sum_{i=1}^{n}\left\{J_{1,n,i}(\theta)-\frac{\gamma\sigma_{0}^{2}}{\sigma^{2}}\partial_{\mu}^{2}b_{i-1}(\mu)Z_{n,i}^{2}B_{i-1}(\mu)\right\}h_{n} \\
&\times{\rm exp}\left(-\frac{\gamma\sigma^{2}_{0}Z_{n,i}^{2}}{2\sigma^{2}}\right)-\gamma\sigma^{-\frac{2+3\gamma}{1+\gamma}}\sum_{i=1}^{n}R_{1,n,i}(\theta), \\
\partial^{2}_{\sigma}Q_{n,\gamma}(\theta)=&-\frac{\gamma(1+2\gamma)}{(1+\gamma)^{2}}\sigma^{-\frac{2+3\gamma}{1+\gamma}}\sum_{i=1}^{n}{\rm exp}\left(-K_{n,i}(\mu,\sigma)-\frac{\gamma\sigma_{0}^{2}Z_{n,i}^{2}}{2\sigma^{2}}\right) \\
&+\frac{2(3+5\gamma)}{1+\gamma}\sigma^{-\frac{2+3\gamma}{1+\gamma}}\sum_{i=1}^{n}{\rm exp}\left(-K_{n,i}(\mu,\sigma)-\frac{\gamma\sigma_{0}^{2}Z_{n,i}^{2}}{2\sigma^{2}}\right)\\
&\times\left(K_{n,i}(\mu,\sigma)+\frac{\gamma\sigma_{0}^{2}Z_{n,i}^{2}}{2\sigma^{2}}\right) \\
&-4\sigma^{-\frac{2+3\gamma}{1+\gamma}}\sum_{i=1}^{n}{\rm exp}\left(-K_{n,i}(\mu,\sigma)-\frac{\gamma\sigma_{0}^{2}Z_{n,i}^{2}}{2\sigma^{2}}\right)\\
&\times\left(K_{n,i}(\mu,\sigma)+\frac{\gamma\sigma_{0}^{2}Z_{n,i}^{2}}{2\sigma^{2}}\right)^{2} \\
=&\sigma^{-\frac{2+3\gamma}{1+\gamma}}\sum_{i=1}^{n}\left(-\frac{\gamma(1+2\gamma)}{(1+\gamma)^{2}}+\frac{\gamma(3+5\gamma)}{1+\gamma}\frac{\sigma_{0}^{2}Z_{n,i}^{2}}{\sigma^{2}}-\frac{\gamma^{2}\sigma_{0}^{4}Z_{n,i}^{4}}{\sigma^{4}}\right)\\
&\times{\rm exp}\left(-\frac{\gamma\sigma_{0}^{2}Z_{n,i}^{2}}{2\sigma^{2}}\right)\\
&+\sigma^{-\frac{2+3\gamma}{1+\gamma}}\sum_{i=1}^{n}\left(\frac{2(3+5\gamma)}{1+\gamma}K_{n,i}(\mu,\sigma)+J_{3,n,i}(\theta)\right){\rm exp}\left(-\frac{\gamma\sigma_{0}^{2}Z_{n,i}^{2}}{2\sigma^{2}}\right) \\
&+\sigma^{-\frac{2+3\gamma}{1+\gamma}}\sum_{i=1}^{n}R_{2,n,i}(\theta){\rm exp}\left(-\frac{\gamma\sigma_{0}^{2}Z_{n,i}^{2}}{2\sigma^{2}}\right)K_{n,i}(\theta)e^{\xi}, 
\end{align*}
\begin{align*}
\partial^{2}_{\mu\sigma}Q_{n,\gamma}(\theta)=&\frac{\gamma^{2}}{1+\gamma}\sigma^{-\frac{2+3\gamma}{1+\gamma}}\sum_{i=1}^{n}{\rm exp}\left(-K_{n,i}(\mu,\sigma)-\frac{\gamma\sigma_{0}^{2}Z_{n,i}^{2}}{2\sigma^{2}}\right)\\
&\times(h_{n}B_{i-1}(\mu)+\sigma_{0}Z_{n,i}\sqrt{h_{n}}+\Delta_{n,i})\partial_{\mu}b_{i-1}(\mu) \\
&+2\gamma\sigma^{-\frac{2+3\gamma}{1+\gamma}}\sum_{i=1}^{n}{\rm exp}\left(-K_{n,i}(\mu,\sigma)-\frac{\gamma\sigma_{0}^{2}Z_{n,i}^{2}}{2\sigma^{2}}\right)\\
&\times(h_{n}B_{i-1}(\mu)+\sigma_{0}Z_{n,i}\sqrt{h_{n}}+\Delta_{n,i})\partial_{\mu}b_{i-1}(\mu)\\
&\times\left(K_{n,i}(\mu,\sigma)+\frac{\gamma\sigma_{0}^{2}Z_{n,i}^{2}}{2\sigma^{2}}\right) \\
&+2\gamma\sigma^{-\frac{2+3\gamma}{1+\gamma}}\sum_{i=1}^{n}{\rm exp}\left(-K_{n,i}(\mu,\sigma)-\frac{\gamma\sigma_{0}^{2}Z_{n,i}^{2}}{2\sigma^{2}}\right)\\
&\times(h_{n}B_{i-1}(\mu)+\sigma_{0}Z_{n,i}\sqrt{h_{n}}+\Delta_{n,i})\partial_{\mu}b_{i-1}(\mu) \\
=&\gamma\sigma^{-\frac{2+3\gamma}{1+\gamma}}\sum_{i=1}^{n}\partial_{\mu}b_{i-1}(\mu)\left(\frac{\gamma}{1+\gamma}+2+\frac{\gamma\sigma_{0}^{2}Z_{n,i}^{2}}{\sigma^{2}}\right)\sigma_{0}Z_{n,i}\sqrt{h_{n}}\\
&\times{\rm exp}\left(-\frac{\gamma\sigma_{0}^{2}Z_{n,i}^{2}}{2\sigma^{2}}\right) \\
&+\gamma\sigma^{-\frac{2+3\gamma}{1+\gamma}}\sum_{i=1}^{n}\partial_{\mu}b_{i-1}(\mu)J_{4,n,i}(\theta){\rm exp}\left(-\frac{\gamma\sigma_{0}^{2}Z_{n,i}^{2}}{2\sigma^{2}}\right) \\
&-\gamma\sigma^{-\frac{2+3\gamma}{1+\gamma}}\sum_{i=1}^{n}\partial_{\mu}b_{i-1}(\mu)R_{3,n,i}(\theta){\rm exp}\left(-\frac{\gamma\sigma_{0}^{2}Z_{n,i}^{2}}{2\sigma^{2}}\right)K_{n,i}(\theta)e^{\xi},
\end{align*}
where $|\xi|\leq |K_{n,i}(\theta)|$ and
\begin{align*}
J_{1,n,i}(\theta)=&B_{i}(\theta)\partial_{\mu}^{2}b_{i-1}(\mu)+\left(\frac{\gamma\sigma_{0}^{2}}{\sigma^{2}}Z_{n,i}^{2}-1\right)\partial_{\mu}b_{i-1}(\mu)\partial_{\mu^{\top}}b_{i-1}(\mu),\\
J_{2,n,i}(\theta)=&\Delta_{n,i}\partial^{2}_{\mu}b_{i-1}(\mu)+\frac{\gamma}{\sigma^{2}}\left\{2\sigma_{0}Z_{n,i}B_{i-1}(\mu)h_{n}^{3/2}+2\sigma_{0}Z_{n,i}\Delta_{n,i}\sqrt{h_{n}}\right.\\
&\left.+B_{i-1}(\mu)^{2}h_{n}^{2}+2B_{i-1}(\mu)\Delta_{n,i}h_{n}+\Delta_{n,i}^{2}\right\}\partial_{\mu}b_{i-1}(\mu)\partial_{\mu^{\top}}b_{i-1}(\mu), \\
J_{3,n,i}(\theta)=&-\frac{4\gamma\sigma_{0}^{2}Z_{n,i}^{2}}{\sigma^{2}}K_{n,i}(\theta)-4K_{n,i}^{2}(\theta) \\
M_{n,i}(\theta)=&\sigma_{0}\partial^{2}_{\mu}b_{i-1}(\mu)Z_{n,i}\sqrt{h_{n}}+J_{1,n,i}(\theta)h_{n}+J_{2,n,i}(\theta) \\
J_{4,n,i}(\theta)=&2\sigma_{0}Z_{n,i}K_{n,i}(\mu,\sigma)\sqrt{h_{n}}+(h_{n}B_{i-1}(\mu)+\Delta_{n,i})\\
&\times\left(\frac{\gamma}{1+\gamma}+2+\frac{\gamma\sigma_{0}^{2}Z_{n,i}^{2}}{\sigma^{2}}+2K_{n,i}(\mu,\sigma)\right) \\
R_{1,n,i}(\theta)=&\{J_{2,n,i}(\theta)-(J_{1,n,i}(\theta)h_{n}+J_{2,n,i}(\theta))\frac{\gamma\sigma_{0}}{\sigma^{2}}Z_{n,i}B_{i-1}(\mu)\sqrt{h_{n}}\}{\rm exp}\left(-\frac{\gamma\sigma^{2}_{0}Z_{n,i}^{2}}{2\sigma^{2}}\right)\\
&+M_{i}(\theta)\left\{\frac{\gamma\sigma_{0}}{\sigma^{2}}Z_{n,i}B_{i-1}(\mu)\sqrt{h_{n}}-K_{n,i}(\theta)+\frac{1}{2}K_{n,i}^{2}(\theta)e^{\xi}\right\}\\
&\times{\rm exp}\left(-\frac{\gamma\sigma^{2}_{0}Z_{n,i}^{2}}{2\sigma^{2}}\right), \\
R_{2,n,i}(\theta)=&-\frac{\gamma(1+2\gamma)}{(1+\gamma)^{2}}+\frac{2(3+5\gamma)}{1+\gamma}\left(\frac{\gamma\sigma^{2}_{0}Z_{n,i}^{2}}{2\sigma^{2}}+K_{n,i}(\theta)\right)-\frac{\gamma^{2}\sigma_{0}^{4}Z_{n,i}^{4}}{\sigma^{4}}+J_{3,n,i}(\theta), \\
R_{3,n,i}(\theta)=&\sigma_{0}\left(\frac{\gamma}{1+\gamma}+2+\frac{\gamma\sigma_{0}^{2}Z_{n,i}^{2}}{\sigma^{2}}\right)Z_{n,i}\sqrt{h_{n}}+J_{4,n,i}(\theta).
\end{align*}
\noindent In the same way as the proof of Theorem 2 in \cite{lee2013minimum}, we can have that
\begin{align*}
\sup_{\theta}\max_{1\leq i\leq n}|J_{2,n,i}(\theta)|&=o_{p}(h_{n}^{r}),\quad 0\leq r<1.5,\\
\sup_{\theta}\max_{1\leq i\leq n}|J_{3,n,i}(\theta)|&=o_{p}(h_{n}^{r}),\quad 0\leq r<0.5,\\
\sup_{\theta}\max_{1\leq i\leq n}|J_{4,n,i}(\theta)|&=o_{p}(h_{n}^{r}),\quad 0\leq r<1,\\
\sup_{\theta}\max_{1\leq i\leq n}|R_{1,n,i}(\theta)|&=o_{p}(h_{n}^{r}),\quad 0\leq r<1.5,\\
\sup_{\theta}\max_{1\leq i\leq n}|R_{2,n,i}(\theta)K_{n,i}(\mu,\sigma)e^{\xi}|&=o_{p}(h_{n}^{r}),\quad 0\leq r<0.5,\\
\sup_{\theta}\max_{1\leq i\leq n}|R_{3,n,i}(\theta)K_{n,i}(\mu,\sigma)e^{\xi}|&=o_{p}(h_{n}^{r}),\quad 0\leq r<1.
\end{align*}

Therefore, we get
\begin{align*}
&\frac{1}{nh_{n}}\partial_{\mu}^{2}Q_{n,\gamma}(\theta)\xrightarrow{p}\gamma\sigma^{-\frac{2+3\gamma}{1+\gamma}}\left(1+\frac{\gamma\sigma^{2}_{0}}{\sigma^{2}}\right)^{-3/2}\\
&\times\left\{S-\int(b(x,\mu_{0})-b(x,\mu))\partial^{2}_{\mu}b(x,\mu)\nu_{0}(x)\right\},\\
&\frac{1}{n}\partial^{2}_{\sigma}Q_{n,\gamma}(\theta)\xrightarrow{p}\sigma^{-\frac{2+3\gamma}{1+\gamma}}\left\{-\frac{\gamma(1+2\gamma)}{(1+\gamma)^{2}}\left(1+\frac{\gamma\sigma_{0}^{2}}{\sigma^{2}}\right)^{-1/2}\right.\\
&\left.+\frac{\gamma(3+5\gamma)}{1+\gamma}\frac{\sigma_{0}^{2}}{\sigma^{2}}\left(1+\frac{\gamma\sigma_{0}^{2}}{\sigma^{2}}\right)^{-3/2}-\frac{3\gamma^{2}\sigma_{0}^{4}}{\sigma^{4}}\left(1+\frac{\gamma\sigma_{0}^{2}}{\sigma^{2}}\right)^{-5/2}\right\}, \\
&\frac{1}{n\sqrt{h_{n}}}\partial_{\mu\sigma}Q_{n,\gamma}(\theta)\xrightarrow{p}0,
\end{align*}
uniformly in $\theta$. This completes the proof of \eqref{eq:C_lim} and therefore Theorem \ref{thm1} is established.

\end{appendices}

%

\bibliographystyle{apalike} 
\bibliography{References.bib}

@article{basu1998robust,
  title={Robust and efficient estimation by minimising a density power divergence},
  author={Basu, Ayanendranath and Harris, Ian R and Hjort, Nils L and Jones, MC},
  journal={Biometrika},
  volume={85},
  number={3},
  pages={549--559},
  year={1998},
  publisher={Oxford University Press}
}

@article{fox1972outliers,
  title={Outliers in time series},
  author={Fox, Anthony J},
  journal={Journal of the Royal Statistical Society: Series B (Methodological)},
  volume={34},
  number={3},
  pages={350--363},
  year={1972},
  publisher={Wiley Online Library}
}

@article{fujisawa2008robust,
  title={Robust parameter estimation with a small bias against heavy contamination},
  author={Fujisawa, Hironori and Eguchi, Shinto},
  journal={Journal of Multivariate Analysis},
  volume={99},
  number={9},
  pages={2053--2081},
  year={2008},
  publisher={Elsevier}
}

@book{hampel1986robust,
  title={Robust statistics},
  author={Hampel, Frank R and Ronchetti, Elvezio M and Rousseeuw, Peter J and Stahel, Werner A},
  year={1986},
  publisher={Wiley Online Library}
}

@book{huber2009robust,
  title={Robust Statistics, Second Edition},
  author={Huber, J and Ronchetti, E. M.},
  year={2009},
  publisher={Wiley}
}

@article{jones2001comparison,
  title={A comparison of related density-based minimum divergence estimators},
  author={Jones, MC and Hjort, Nils Lid and Harris, Ian R and Basu, Ayanendranath},
  journal={Biometrika},
  volume={88},
  number={3},
  pages={865--873},
  year={2001},
  publisher={Oxford University Press}
}

@article{kessler1997estimation,
  title={Estimation of an ergodic diffusion from discrete observations},
  author={Kessler, Mathieu},
  journal={Scandinavian Journal of Statistics},
  volume={24},
  number={2},
  pages={211--229},
  year={1997},
  publisher={Wiley Online Library}
}

@article{lee2013minimum,
  title={Minimum density power divergence estimator for diffusion processes},
  author={Lee, Sangyeol and Song, Junmo},
  journal={Annals of the Institute of Statistical Mathematics},
  volume={65},
  number={2},
  pages={213--236},
  year={2013},
  publisher={Springer}
}

@book{maronna2019robust,
  title={Robust statistics: theory and methods (with R)},
  author={Maronna, Ricardo A and Martin, R Douglas and Yohai, Victor J and Salibi{\'a}n-Barrera, Mat{\'\i}as},
  year={2019},
  publisher={John Wiley \& Sons}
}

@article{uchida2012adaptive,
  title={Adaptive estimation of an ergodic diffusion process based on sampled data},
  author={Uchida, Masayuki and Yoshida, Nakahiro},
  journal={Stochastic Processes and their Applications},
  volume={122},
  number={8},
  pages={2885--2924},
  year={2012},
  publisher={Elsevier}
}

@article{yoshida2011polynomial,
  title={Polynomial type large deviation inequalities and quasi-likelihood analysis for stochastic differential equations},
  author={Yoshida, Nakahiro},
  journal={Annals of the Institute of Statistical Mathematics},
  volume={63},
  number={3},
  pages={431--479},
  year={2011},
  publisher={Springer}
}

@article{la2010infinitesimal,
  title={Infinitesimal robustness for diffusions},
  author={La Vecchia, Davide and Trojani, Fabio},
  journal={Journal of the American Statistical Association},
  volume={105},
  number={490},
  pages={703--712},
  year={2010},
  publisher={Taylor \& Francis}
}

@article{song2017robust,
  title={Robust estimation of dispersion parameter in discretely observed diffusion processes},
  author={Song, Junmo},
  journal={Statistica Sinica},
  pages={373--388},
  year={2017},
  publisher={JSTOR}
}

@article{ghosh2022general,
  title={GENERAL ROBUST BAYES PSEUDO-POSTERIORS: EXPONENTIAL CONVERGENCE RESULTS WITH APPLICATIONS},
  author={Ghosh, Abhik and Majumder, Tuhin and Basu, Ayanendranath},
  journal={Statistica Sinica},
  volume={32},
  pages={787--823},
  year={2022}
}

@article{momozaki2022robustness,
  title={Robustness against outliers in ordinal response model via divergence approach},
  author={Momozaki, Tomotaka and Nakagawa, Tomoyuki},
  journal={arXiv preprint arXiv:2209.11965},
  year={2022}
}

@article{song2007minimum,
  title={Minimum density power divergence estimator for diffusion parameter in discretely observed diffusion processes},
  author={Song, Jun-Mo and Lee, Sang-Yeol and Na, Ok-Young and Kim, Hyo-Jung},
  journal={Communications for Statistical Applications and Methods},
  volume={14},
  number={2},
  pages={267--280},
  year={2007},
  publisher={The Korean Statistical Society}
}

@book{pardo2006statistical,
  title={Statistical inference based on divergence measures},
  author={Pardo, Leandro},
  year={2006},
  publisher={CRC press}
}

@article{cichocki2010families,
  title={Families of alpha-beta-and gamma-divergences: Flexible and robust measures of similarities},
  author={Cichocki, Andrzej and Amari, Shunichi},
  journal={Entropy},
  volume={12},
  number={6},
  pages={1532--1568},
  year={2010},
  publisher={MDPI}
}

@article{yoshida1988robust,
  title={Robust {M}-estimators in diffusion processes},
  author={Yoshida, Nakahiro},
  journal={Annals of the Institute of Statistical Mathematics},
  volume={40},
  pages={799--820},
  year={1988},
  publisher={Kluwer Academic Publishers-Plenum Publishers}
}

@article{masuda2017moment,
  title={Moment convergence in regularized estimation under multiple and mixed-rates asymptotics},
  author={Masuda, Hiroki and Shimizu, Yusuke},
  journal={Mathematical Methods of Statistics},
  volume={26},
  pages={81--110},
  year={2017},
  publisher={Springer}
}
\end{document}